\documentclass[11pt]{iopart}
\expandafter\let\csname equation*\endcsname\relax
\expandafter\let\csname endequation*\endcsname\relax
\usepackage[T1]{fontenc}
\usepackage[utf8]{inputenc}
\usepackage{amsmath}
\usepackage{graphicx,bbm,amssymb,amsopn}
\newcommand\scalemath[2]{\scalebox{#1}{\mbox{\ensuremath{\displaystyle #2}}}}

\newcommand{\bra}[1]{{\langle #1 \vert}}

\newcommand{\ket}[1]{{\vert #1 \rangle}}

\newcommand{\ii}{ {\rm i} }

\newcommand{\ZZ}{\mathbb{Z}}
\newcommand{\RR}{\mathbb{R}}
\newcommand{\CC}{\mathbb{C}}

\newcommand{\be}{\begin{equation}}
\newcommand{\ee}{\end{equation}}
\newcommand{\bea}{\begin{eqnarray}}
\newcommand{\eea}{\end{eqnarray}}

\newcommand{\hm}[2]{{\hat{#1}^{(#2)}}}
\newcommand{\hmp}[2]{{\hat{#1}^{'(#2)}}}
\newcommand{\hmu}[2]{{\hat{\mathbf{#1}}_{#2}}}
\newcommand{\hmup}[2]{{\hat{\mathbf{#1}}'_{#2}}}
\newcommand{\mmu}[2]{{\mathbf{#1}_{#2}}}
\newcommand{\mmup}[2]{{\mathbf{#1}'_{#2}}}

\def\one{\mathbbm{1}}

\def\WW{{{\hat{\mathbf{W}}}}}

\begin{document}

\title[Exact matrix product decay modes of a boundary driven cellular automaton]{Exact matrix product decay modes of a boundary driven cellular automaton}

\author{Toma\v{z} Prosen$^1$ and Berislav Bu\v ca$^2$}
\address{$^1$Department of Physics, Faculty of Mathematics and Physics, University of Ljubljana, Ljubljana, Slovenia}
\address{$^2$Department of Medical Physics and Biophysics, University of Split School of Medicine, Split, Croatia}

\begin{abstract}
We study integrability properties of a reversible deterministic cellular automaton (the rule 54 of [Boben\-ko {\em et al}, Commun. Math. Phys. {\bf 158}, 127 (1993)]) and present a bulk algebraic relation and its inhomogeneous extension which allow for an explicit construction of Liouvillian decay modes for two distinct families of stochastic boundary driving. The spectrum of the many-body stochastic matrix defining the time propagation is found to separate into sets, which we call \emph{orbitals}, and the eigenvalues in each orbital are found to obey a distinct set of Bethe-like equations. We construct the decay modes in the first orbital (containing the leading decay mode) in terms of an exact inhomogeneous matrix product ansatz, study the thermodynamic properties of the spectrum and the scaling of its gap, and provide a conjecture for the Bethe-like equations for all the orbitals and their degeneracy.   
\end{abstract}

\section{Introduction}

Understanding the emergence of laws governing macroscopic physical phenomena, such as transport and relaxation, from deterministic and reversible microscopic dynamics is one of the most prominent fundamental problems of statistical mechanics. In this context, an important setup consists of driving a finite many-body system, say a one dimensional lattice with local interactions, with a pair of macroscopic (infinite) reservoirs attached, or coupled to the system's ends (boundaries). Infinite reservoirs can typically be replaced by stochastic forces acting on boundary degrees of freedom of the system, so we are speaking of {\em boundary driven deterministic dynamics}. Several non-trivial exactly solvable examples of current carrying non-equilibrium steady states (NESS) of this type of dynamics have been recently found in the realm integrable quantum lattice models \cite{review}, however all attempts of exact constructions of dynamical modes of relaxation have failed so far.

Recently, NESS of a boundary driven reversible cellular automaton, which can be understood as a simple caricature of deterministic interacting dynamics, has been found \cite{drivenbobenko} and its construction exhibits certain interesting algebraic properties. The cellular automaton is the Rule 54 of Bobenko {\em et al} \cite{Bobenko}, which is a two-state fully deterministic, reversible many-body interacting system and admits non-trivially scattering solitons. 

\begin{figure}
 \centering	
\vspace{-1mm}
\includegraphics[width=\columnwidth]{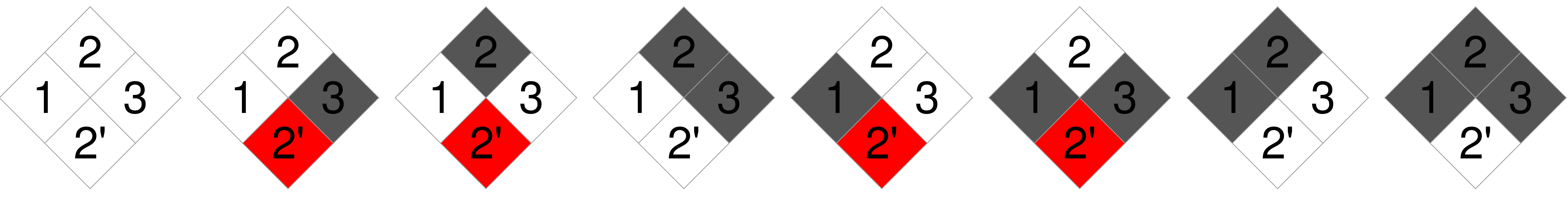}
\vspace{-1mm}
\caption{The local rule 54. Each site can be either in state 0 or 1. State 1 (occupied) is shown as dark gray (at the current time step) and as red (at the next time step), state 0 (empty) is white. The state of site 2 at the next time step (denoted with 2') is determined by the state of sites 1, 2, and 3 at the current time step. By combining these plaquettes along the rows of a diamond lattice one builds the time evolution of the model. 
It can easily be seen that this leads to solitons (all traveling at the same velocity) which can scatter and incur a shift. These are the elementary modes of the model.}
\label{fig1}
\end{figure}

The rule is given by a deterministic local mapping on a diamond-shaped plaquette $\chi : \ZZ_2\times \ZZ_2\times \ZZ_2 \to \ZZ_2$. A south site $s_{\rm S}$ is determined by a north, west and east sites
\be
s_{\rm S} = \chi({s_{\rm W},s_{\rm N},s_{\rm E}}) = s_{\rm N} + s_{\rm W} + s_{\rm E} +  s_{\rm W} s_{\rm E} \pmod{2}, \quad s_{\rm S},s_{\rm N},s_{\rm W},s_{\rm E} \in \ZZ_2.\label{chi}
\ee
Time runs in the north to south direction (see Fig.~\ref{fig1}) and defines a simple interacting dynamics over a $1+1$ dimensional lattice $s_{x,t+1} = \chi(s_{x-1,t},s_{x,t-1},s_{x+1,t})$, where only lattice sites $(x,t)$ of fixed parity of $x\pm t$ are considered.

\begin{figure}
 \centering	
\vspace{-1mm}
\includegraphics[width=0.6\columnwidth]{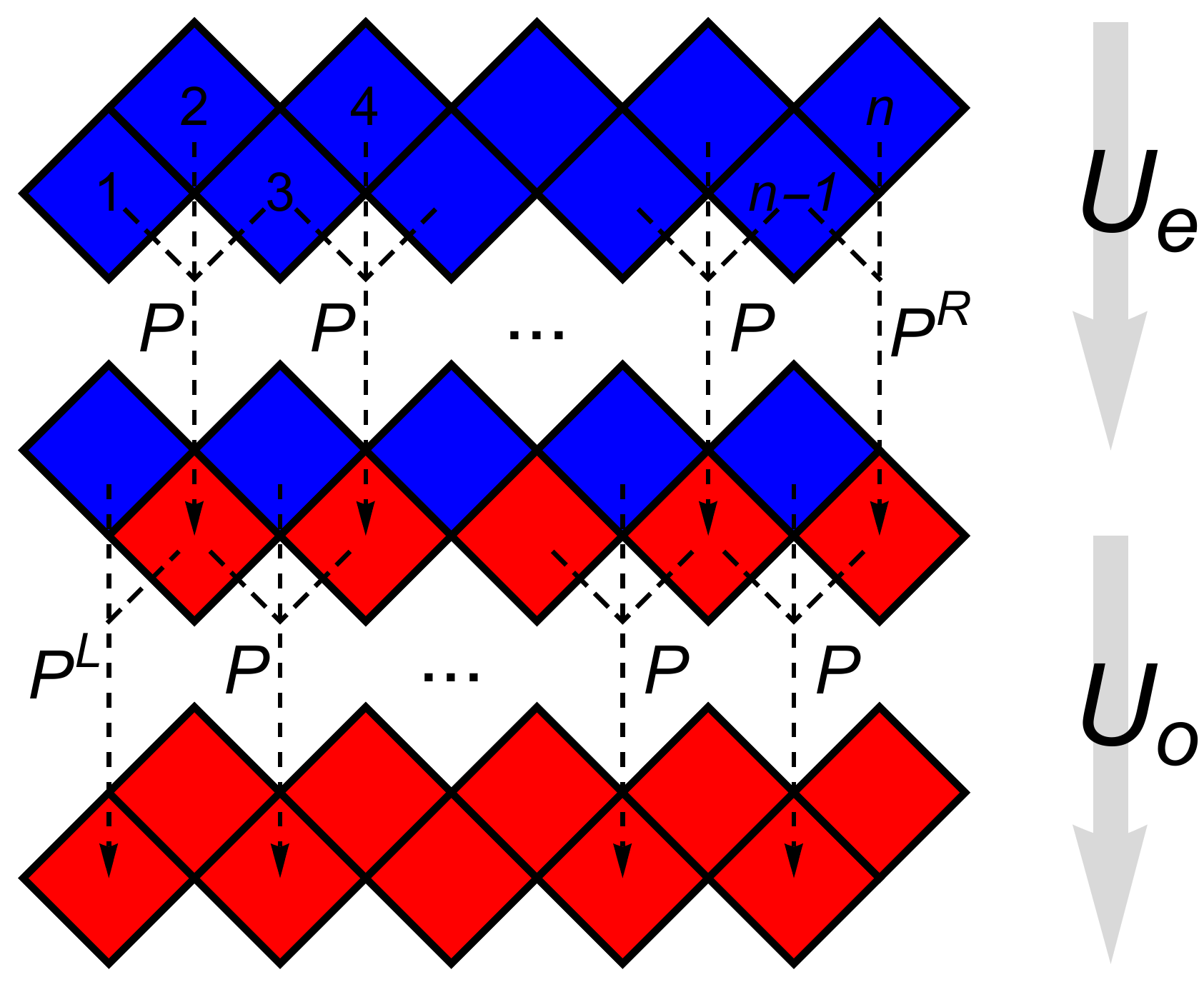}
\vspace{-1mm}
\caption{Schematic construction of the propagator (\ref{U}), composed of $U_{\rm e}$ and $U_{\rm o}$, where each time layer is in turn composed of mutually commuting three site local permutation maps $P$ and two site boundary stochastic maps $P^{\rm L,R}$ (see Eqs.~(\ref{Ue},\ref{Uo})). In blue/red we denote the sites before/after the update.}
\label{fig2}
\end{figure}

We shall now define dynamics over a finite chain of even number of sites $n$ with the initial data given by a configuration along a saw $(s_1,s_2,\ldots,s_n) \equiv (s_{1,t+1},s_{2,t},s_{3,t+1},s_{4,t},\ldots,s_{n-1,t+1},s_{n,t})$ (see Fig.~\ref{fig2}), which can be given as a composition of even site updates $s_{2 y,t+1}=\chi(s_{2y-1,t},s_{2y,t-1},s_{2y+1,t})$ and odd site updates $s_{2y+1,t+2}=\chi(s_{2y,t+1},s_{2y+1,t},s_{2y+2,t+1})$.
The dynamics is fully deterministic, except for the sites near the boundaries, where we shall prescribe appropriate local Markov stochastic processes by which we drive the model out of equilibrium.

We thus proceed to formulate the time evolution of the full probability distribution $p_{(s_1,s_2,\ldots,s_n)}$, which we call  a {\em state}.\footnote{This notion of the state (as a macro-state) should be distinguished from a binary state of an automaton. Since the meaning of the term should be clear from the context we use the same word for the two concepts.}
The state space is a convex subset of a vector space ${\cal S} = \RR^{\cal C} = (\RR^2)^{\otimes n}$ of probability distributions over configurations $\mathbf{p}=(p_0,p_1,\ldots,p_{2^n-1}) \in {\cal S}$ satisfying the non-negativity and normalization conditions,
$p_s \ge 0$, $\sum_{s=0}^{2^n-1} p_s = 1$. Here, $s$ is a binary coded configuration $s=\sum_{k=1}^n 2^{n-k} s_k$. 
The local rule 54 can then be given in terms of a three-site $2^3 \times 2^3$ permutation matrix $P$
\begin{equation}
P_{(s,s',s''),(t,t',t'')} = \delta_{s,t} \delta_{s',\chi(t,t',t'')} \delta_{s'',t''},
\end{equation}
or
$$
P = \begin{pmatrix} 
\, 1\, & & & & & & & \cr
& & & \, 1\, & & & & \cr
& & \, 1\, & & & & & \cr
& \,1\, & & & & & & \cr
& & & & & & \,1\, & \cr
& & & & & & & \,1\, \cr
& & & & \,1\, & & & \cr
& & & & & \,1\, & & \end{pmatrix},
$$
such that $P^2=\one$. The local update rule is in turn embedded into ${\rm End}({\cal S})$ as $P_{k,k+1,k+2} = \one_{2^{k-1}}\otimes P \otimes \one_{2^{n-k-2}}$ acting on any triple of neighboring sites $k, k+1, k+2$.
The time evolution of the state vector $\mathbf{p}(t) \in {\cal S}$ starting from some initial state $\mathbf{p}(0)$ is written as
\begin{equation}
\mathbf{p}(t)=U^t \mathbf{p}(0),
\end{equation}
where 
\be
U=U_{\rm o} U_{\rm e}, \label{U}
\ee
is the one-step propagator that is factored in terms of two temporal layers which generate staggered dynamics for, respectively, even and odd sites
\begin{align}
&U_{\rm e}=P_{123}P_{345}\cdots P_{n-3,n-2,n-1} P^{\rm{R}}_{n-1,n}, \label{Ue}\\
&U_{\rm o}= P^{\rm{L}}_{12} P_{234} P_{456}  \cdots P_{n-2,n-1,n}. \label{Uo}
\end{align}
The boundary propagators
\be
P^{\rm{L}}_{12}=P^{\rm{L}} \otimes \one_{2^{n-2}}, \qquad P^{\rm{R}}_{n-1,n}= \one_{2^{n-2}} \otimes P^{\rm{R}}.
\ee
are given in terms of $4\times 4$ stochastic matrices\footnote{By definition, stochastic matrices have non-negative elements which in each column sum to 1.}
 $P^{\rm L}$ and $P^{\rm R}$ (to be specified later), which in turn imply that the full $2^n\times 2^n$ propagator $U$ itself is a stochastic matrix 
 and thus conserves total probability during the time evolution. This dynamics which is bulk-deterministic and boundary-stochastic should be contrasted with related, though distinct, 
 discrete time asymmetric exclusion process models \cite{asep1, asep2, asep3}, which feature both stochastic bulk dynamics as well as stochastic driving.
 
We note that $P_{k-1,k,k+1}$ changes only the site $k$, conditioned on the states of the sites $k-1$ and $k+1$, so the local propagators commute if at least two sites apart
\be
[P_{k-1,k,k+1},P_{k'-1,k',k'+1}] = 0,\quad {\rm if}\quad |k-k'| \ge 2.
\ee
Furthermore we shall require that also boundary stochastic matrices commute with the neighbouring bulk propagators
\be
[P^{\rm L}_{12},P_{234}]=0,\qquad
[P_{123},P^{\rm R}_{34}]=0,
 \label{boundcomm}
\ee
so the order of factors in either $U_{\rm e}$ or $U_{\rm o}$, Eqs. (\ref{Ue},\ref{Uo}), is irrelevant.

The main objective of this paper is an exact solution of an eigenvalue equation for the Markov propagator
\begin{equation}
U \mathbf{p}=\Lambda \mathbf{p},
\end{equation}
which can be conveniently split into a pair of equations for an eigenstate at even and odd time layers with the eigenvalue $\Lambda$ factored into left and right parts,
\begin{equation}
U_{\rm e} \mathbf{p}=\Lambda_{\rm{L}} \mathbf{p'}, \qquad U_{\rm o} \mathbf{p'}=\Lambda_{\rm{R}} \mathbf{p}, \qquad \Lambda=\Lambda_{\rm{R}} \Lambda_{\rm{L}}. \label{eigensplit}
\end{equation}
As shown in Ref.~\cite{drivenbobenko} (Theorem 1) for boundary stochastic matrices, having nonvanishing both rates for stochastically setting the state of the site near the boundary,
the propagator $U$ is {\em irreducible} and {\em aperiodic}. Hence, according to Perron-Frobenius theorem, the non-equilibrium steady state (NESS) eigenvector, corresponding to $\Lambda_0=1$, is unique and  all other eigenvalues $\Lambda_{j\ge 1}$ are bounded by $|\Lambda_j|<1$ and thus the corresponding components of the state vector decay during the time evolution. These eigenvectors are also called decay modes, as they encode time evolution of any state as
\be
\mathbf{p}(t) = \mathbf{p}_0 + \sum_{j\ge 1} c_j \Lambda_j^t \mathbf{p}_j,
\ee 
where $c_j$ are appropriate constants depending on the initial state. 

In this paper we formulate a compact matrix product ansatz (MPA) which encodes eigenvectors of $U$ for the most important part of the spectrum, including NESS and the
leading decay mode determining the spectral gap of $U$. In particular, we find that the spectrum organizes into \emph{orbitals}, i.e., the sets of eigenvalues fulfilling the same Bethe equations (see Fig.~\ref{fig:orbitals} in Sec.~\ref{subsec:thermo}). The NESS belongs to an especially simple (zeroth) orbital that contains also three additional eigenvectors whose eigenvalues do not depend on the system size. The first orbital contains the leading decay mode. The eigenvalues of the zeroth and the first orbital are nondegenerate. The other orbitals seem to be exponentially degenerate in system size. 

We consider two types of boundary driving for which exact solutions can be found, the first we call \emph{conditional driving}, and the second \emph{Bernoulli driving}. Bernoulli driving has been introduced and studied for the steady state in Ref.~\cite{drivenbobenko}. 

\subsection{Conditional boundary driving}

For the conditional driving the boundary stochastic matrices read 
\begin{align}
&P^{\rm{L}}=\left(
\begin{array}{cccc}
 \alpha  & 0 & \alpha  & 0 \\
 0 & \beta  & 0 & \beta  \\
 1-\alpha  & 0 & 1-\alpha  & 0 \\
 0 & 1-\beta  & 0 & 1-\beta  \\
\end{array}
\right), \quad P^{\rm{R}}=\left(
\begin{array}{cccc}
 \gamma  & \gamma  & 0 & 0 \\
 1-\gamma  & 1-\gamma  & 0 & 0 \\
 0 & 0 & \delta  & \delta  \\
 0 & 0 & 1-\delta  & 1-\delta  \\
\end{array}
\right), 
\label{cond}
\end{align}
where $\alpha, \beta, \gamma, \delta \in [0,1]$ are some driving rates parametrizing the left and the right bath. We call this conditional driving since in $P^{\rm{L}}_{12}$ ($P^{\rm{R}}_{n-1,n}$) the probability of changing the site 1 depends {\em only} on the state of the neighboring site 2 (changing the site $n$ depends only on the state of the site $n-1$). For instance, if the site 2 is in state 0 then the site 1 will be stochastically set to state 0 with the rate $\alpha$ or to state 1 with the rate $1-\alpha$. If on the other hand, the site 2 is in state 1, the site 1 will be set to state 0 or 1 with the rates $\beta$ or $1-\beta$. The analogous also holds for $P^{\rm{R}}_{n-1,n}$.

\subsection{Bernoulli boundary driving}

For the Bernoulli driving, explained in more detail in Ref.~\cite{drivenbobenko}, we have
\begin{align}
&P^{\rm{L}}=\left(
\begin{array}{cccc}
 \frac{1}{2} & 0 & \frac{1}{2} & 0 \\
 0 & 1-\alpha  & 0 & \beta  \\
 \frac{1}{2} & 0 & \frac{1}{2} & 0 \\
 0 & \alpha  & 0 & 1-\beta  \\
\end{array}
\right), \qquad P^{\rm{R}}=\left(
\begin{array}{cccc}
 \frac{1}{2} & \frac{1}{2} & 0 & 0 \\
 \frac{1}{2} & \frac{1}{2} & 0 & 0 \\
 0 & 0 & 1-\gamma  & \delta  \\
 0 & 0 & \gamma  & 1-\delta  \\
\end{array}
\right),  \label{bern}
\end{align}
where $\alpha, \beta, \gamma, \delta \in [0,1]$ are again some driving rates specifying the left and the right bath. However, for this case it turns out that all the results are more compactly and conveniently expressed in terms of another set of, so called, {\em difference parameters}
\begin{equation}
\mu:=\alpha-\beta, \qquad \nu:=\beta-1, \qquad \sigma:=\gamma-\delta, \qquad \rho:=\delta-1. \nonumber 
\end{equation}
We note that both sets of boundary driving stochastic matrices (\ref{cond},\ref{bern}) satisfy the commutativity condition (\ref{boundcomm}) and represent so far the only known exactly solvable boundaries for the 
Rule 54. The converse seems not to be true. Not any pair of stochastic boundary matrices which satisfy (\ref{boundcomm}) allow for exact solutions. Note as well that both type of boundary matrices (\ref{cond},\ref{bern}) satisfy the conditions of 
`holographic ergodicity' theorem of \cite{drivenbobenko}, implying exponential decay of any initial state to a unique NESS, for an open set of parameters $0 < \alpha,\beta,\gamma,\delta < 1$.
\\\\
The rest of the paper is organized as follows. In Sec.~\ref{sec:NESS} we introduce a cubic bulk algebra, which seems to provide the fundamental integrability relation of the model, and use it to solve the NESS-orbital in terms of MPA for both drivings (the NESS for the Bernoulli driving case was solved in terms of an alternative, patch-state ansatz in Ref.~\cite{drivenbobenko}). In Sec.~\ref{sec:firstorbital} we introduce a generalization of the aforementioned algebra and use it to construct eigenvectors in the first orbital in terms of an {\em inhomogeneous} (spatially modulated) MPA. This orbital also contains the leading decay mode. The consistency conditions lead to a simple set of Bethe-like equations which yield the spectrum of the first orbital.
We also discuss the thermodynamic limit and show that the spectral gap closes as $~1/n$. We close the section by discussing how the cubic bulk algebra may be re-written as a quadratic algebra, somewhat similar to Zamolodchikov-Faddeev (ZF) algebra.
In Sec. \ref{sec:conjectures} we provide a conjecture for the Bethe-like equations for the entire spectrum, as well as a conjecture for the degeneracy of the higher orbital eigenstates. 
Finally we end with the conclusions. The paper also contains an  \ref{boundaryvecs} stating explicitly all the components of the boundary vectors of the MPA generating the decay modes of the first orbital,.    

Throughout the paper, whenever we provide explicit results, we will first state the results for the conditional driving \eqref{cond} and then for the Bernoulli driving \eqref{bern}.       

\section{The cubic algebra and the non-equilibrium steady state}
\label{sec:NESS}
Let us first fix some notation. Quantities that are vectors in the physical space are written in bold-face. The numeral subscript of a physical space vector (or operator) denotes the site position in the tensor product $(\RR^2)^{\otimes n}$. When writing in component notation the components in physical space are labeled with a binary `spin' index, such as $s\in\{0,1\}$. Matrices are written with capital roman letters. These are typically acting over 4-dimensional auxiliary space, except for the propagator $U$ and its local pieces 
$P_{k-1,k,k+1}, P^{\rm{L}}_{12}, P^{\rm{R}}_{n-1,n}$ which are operators in the physical space and act trivially in the auxiliary space. Matrices in an extended 8-dimensional auxiliary space (employed in the next section) will be denoted with hats. Row (column) vectors in the auxiliary space will be written as Dirac bras (kets). 

We begin by defining a vector
\be
\mathbf{W}= \begin{pmatrix}
W_0(\xi, \omega)\cr 
W_1(\xi, \omega)
\end{pmatrix},
\ee
with components $W_s$ being $4\times 4$ matrices
\be
W_0(\xi, \omega)=\left(
\begin{array}{cccc}
 1 & 1 & 0 & 0 \\
 0 & 0 & 0 & 0 \\
 \xi  & \xi  & 0 & 0 \\
 0 & 0 & 0 & 0 \\
\end{array}
\right), \qquad 
W_1(\xi, \omega)=\left(
\begin{array}{cccc}
 0 & 0 & 0 & 0 \\
 0 & 0 & \xi  & 1 \\
 0 & 0 & 0 & 0 \\
 0 & 0 & 1 & \omega  \\
\end{array}
\right),
\label{Wrepr}
\ee
depending on a pair of formal parameters $\xi$ and $\omega$, which we call {\em spectral parameters}. 
We also define a matrix $\mathbf{W}'$, which is $\mathbf{W}$ with $\xi$ and $\omega$ interchanged, i.e., (writing the dependence on the spectral parameters explicitly) 
\be
\mathbf{W}'(\xi, \omega)=\mathbf{W}(\omega,\xi).
\ee

The key property explored in this paper is a simple three-site cubic algebraic relation\footnote{Note a similar two-site cancellation mechanism in discrete time ASEP models \cite{asep1,asep2,asep3}.} which shall provide a cancellation mechanism to be used later for constructing the eigenstates of the Markov matrix
\be
P_{123} \mathbf{W}_1 S\, \mathbf{W}_2 \mathbf{W}'_3=\mathbf{W}_1 \mathbf{W}'_2 \mathbf{W}_3 S,  \label{bulk1}
\ee
where $S$ is a constant `delimiter' matrix, 
\be
S=\one \otimes \sigma^{\rm{x}}=\left(
\begin{array}{cccc}
 0 & 1 & 0 & 0 \\
 1 & 0 & 0 & 0 \\
 0 & 0 & 0 & 1 \\
 0 & 0 & 1 & 0 \\
\end{array}
\right),
\label{Sdef}
\ee
satisfying $S^2=\one_4$.
By interchanging $\xi$ and $\omega$ in \eqref{bulk1}, i.e. interchanging $\mathbf{W}$ and $\mathbf{W}'$, and multiplying with $P_{123}$ (noting that $P^2=\one_8$) we obtain a
dual bulk relation
\be
P_{123} \mathbf{W}'_1 \mathbf{W}_2 \mathbf{W}'_3 S = \mathbf{W}'_1 S\, \mathbf{W}'_2 \mathbf{W}_3.  \label{bulk1b}
\ee
Note that Eq.~(\ref{bulk1}) in fact represents $8$ matrix product identities,
$W_s S W_{\chi(ss's'')} W'_{s''} = W_s W'_{s'} W_{s''} S$, and analogously for Eq.~$(\ref{bulk1b})$, by writing out physical space components $s,s',s''\in\{0,1\}$ for a vector on a triple of consecutive physical sites (denoted in (\ref{bulk1},\ref{bulk1b}) as $123$). Eqs. (\ref{bulk1},\ref{bulk1b}) thus represent a set of algebraic relations among $W_0,W_1,W'_0,W'_1,S$ which can be straightforwardly verified for the representation \eqref{Wrepr}.

We shall begin by proposing a simple ansatz for the eigenvectors of $U$ in terms of the following staggered matrix product states
\begin{align}
&\mathbf{p}=\bra{\mathbf{l}_1}\mmu{W}{2}\mmup{W}{3} \mmu{W}{4} \mmup{W}{5}\cdots\mmup{W}{n-3}\mmu{W}{n-2}\ket{\mathbf{r}_{n-1,n}}, \label{pness} \\
&\mathbf{p'}=\bra{\mathbf{l}'_{12}}\mmu{W}{3}\mmup{W}4\mmu{W}{5}\mmup{W}{6}\cdots\mmup{W}{n-2}\mmu{W}{n-1}\ket{\mathbf{r}'_{n}}. \label{ppness}
\end{align}
In order for fixed point condition \eqref{eigensplit} to hold for $\mathbf{p},\mathbf{p'}$ we require the following boundary conditions to be satisfied
\begin{align}
&P^{\rm{L}}_{12} \bra{\mathbf{l}'_{12}}=\lambda_{\rm{L}} \bra{\mathbf{l}_1} \mmu{W}{2} S, \label{bound1a} \\
&P_{123} \bra{\mathbf{l}_{1}} \mmu{W}{2} \mmup{W}{3}=\bra{\mathbf{l}'_{12}} \mmu{W}{3}S,\label{bound2a} \\
&P^{R}_{12} \ket{\mathbf{r}_{12}}=\mmup{W}{1}S\ket{\mathbf{r}'_2}, \label{bound3a} \\
&P_{123} \mmup{W}{1} \mmu{W}{2}\ket{\mathbf{r}'_{3}} =\lambda_{\rm{R}} \mmup{W}{1}S\ket{\mathbf{r}_{23}}. \label{bound4a}
\end{align}
Specifically, writing out $U_{\rm e} \mathbf{p}$ in terms of (\ref{Ue}) and the ansatz \eqref{pness}, one first uses Eq.~\eqref{bound3a} in order to introduce the delimiter $S$ in a string 
$\cdots \mathbf{W}'_{n-3}\mathbf{W}_{n-2}\mathbf{W}'_{n-1}S$ and then implements $\cdots P_{n-5,n-4,n-3}P_{n-3,n-2,n-1}$ via the dual bulk relation \eqref{bulk1b} in order to move the delimiter $S$ across to the left end where it is then absorbed, via $S^2=\one$, by applying another boundary equation \eqref{bound2a}, arriving to $U_{\rm e} \mathbf{p}=\lambda_{\rm R}\mathbf{p}'$.
Analogously we proceed with $U_{\rm o} \mathbf{p}'$, in terms of (\ref{Uo}) and ansatz \eqref{ppness}, now implementing the boundary Eqs.~(\ref{bound1a},\ref{bound4a}) to carry $S$
from left to right via the bulk relation \eqref{bulk1}, ending with $U_{\rm e} \mathbf{p}'=\lambda_{\rm L}\mathbf{p}$. Thus, $\mathbf{p}$ is an eigenvector of $U$ with an eigenvalue\footnote{
Please note the use of small letters $\lambda,\lambda_{\rm L,R}$ for designating the spectral variables for the NESS-orbital in distinction to capitalised variables $\Lambda,\Lambda_{\rm L,R}$ referring to the
general case \eqref{eigensplit} which, in the case of the first orbital, can be expressed as functions of the NESS-orbital data (see Sec.~\ref{sec:firstorbital}).}
$\lambda=\lambda_{\rm L}\lambda_{\rm R}$.

Solving the full set of boundary equations (\ref{bound1a}-\ref{bound4a}) should fix all the unknown parameters in the MPA (\ref{pness},\ref{ppness}) as the bulk relations are automatically satisfied. The solution is unique up to an irrelevant transformation of boundary vectors. We first focus on the conditional driving case \eqref{cond} and then state the results for Bernoulli driving \eqref{bern}. 

Solving separately the pair of boundary equations for the left side (\ref{bound1a},\ref{bound2a}) we obtain the following unique solutions for the spectral parameters,
\begin{align}
&\xi=\frac{(\alpha +\beta -1)-\beta  \lambda_{\rm{L}}}{(\beta -1) \lambda_{L}^{2}}, \label{nessspc1} \\
&\omega =\frac{\lambda_{\rm{L}} (\alpha -\lambda_{\rm{L}})}{(\beta -1)}, \label{nessspc2}
\end{align}
and for the left boundary vectors (in physical space components)
\begin{align}
&\bra{l_0}=\left( 0, \frac{1}{\lambda_{\rm{L}}}, \frac{\alpha  (\alpha +\beta -1-\lambda_{\rm{L}})}{\beta  \lambda_{\rm{L}}^2 (\alpha -\lambda_{\rm{L}})},1 \right), \nonumber \\
&\bra{l_1}= \left(0, \frac{1-\beta }{\beta  \lambda_{\rm{L}}}, \frac{(1-\alpha ) (\alpha +\beta-1-\lambda_{\rm{L}})}{\beta  \lambda_{\rm{L}}^2 (\alpha -\lambda_{\rm{L}})},\frac{1}{\beta }-1\right), \nonumber \\
&\bra{l'_{0,0}}=\left(\frac{\alpha  (\alpha +\beta -1-\lambda_{\rm{L}})}{(\beta -1) \beta  \lambda_{\rm{L}}}, \frac{\alpha +\beta -1-\lambda_{\rm{L}}}{\beta -1}, 0, 0  \right), \nonumber \\
&\bra{l'_{0,1}}=\left( 0, 0, \frac{\lambda_{\rm{L}} (\alpha -\lambda_{\rm{L}}) (\alpha +\beta-1-\lambda_{\rm{L}})}{(\beta -1) (\alpha+\beta -1 -\beta  \lambda_{\rm{L}})},\frac{\alpha  (\alpha +\beta -1-\lambda_{\rm{L}})}{(\beta -1) \beta \lambda_{\rm{L}}} \right), \nonumber \\
&\bra{l'_{1,0}}=\left( \frac{(\alpha -\lambda_{\rm{L}}) (\alpha +\beta-1 -\lambda_{\rm{L}})}{(1-\beta ) \beta  \lambda_{\rm{L}}}, -\frac{\alpha +\beta-1 -\lambda_{\rm{L}}}{\beta }, 0, 0 \right), \nonumber \\
&\bra{l'_{1,1}}= \left( 0, 0, \frac{(1-\alpha ) \lambda_{\rm{L}} (\alpha +\beta -1-\lambda_{\rm{L}})}{\beta  (\alpha+\beta-1 -\beta \lambda_{\rm{L}})},\frac{(1-\alpha )(\alpha +\beta-1 -\lambda_{\rm{L}})}{(\beta -1) \beta  \lambda_{\rm{L}}}  \right). 
\end{align}
The right boundary equations (\ref{bound3a},\ref{bound4a}), on the other hand, give the following unique solutions for the spectral parameters
\begin{align}
&\xi=\frac{\lambda_{\rm{R}} (\gamma -\lambda_{\rm{R}})}{(\delta -1) }, \label{nessspc3} \\
&\omega =\frac{ (\gamma +\delta -1)-\delta  \lambda_{\rm{R}}}{(\delta -1) \lambda_{\rm{R}}^2},\label{nessspc4}
\end{align}
and the the right boundary vectors 
\begin{align}
&\ket{r_{0,0}}=\left( \frac{\gamma  (\gamma +\delta-1 -\lambda_{\rm{R}})}{\delta  (\gamma +\delta -1
   -\delta \lambda_{\rm{R}})}, 0, \frac{\gamma +\delta-1 -\lambda_{\rm{R}}}{(\delta -1) \lambda_{\rm{R}}}, 0 \right), \nonumber \\   
&\ket{r_{0,1}} =\left( \frac{(\lambda_{\rm{R}}-\gamma ) (\gamma +\delta-1 -\lambda_{\rm{R}})}{\delta (\gamma +\delta-1 - \delta \lambda_{\rm{R}} )}, 0,- \frac{\gamma +\delta-1 -\lambda_{\rm{R}}}{\delta  \lambda_{\rm{R}}} , 0 \right), \nonumber \\
&\ket{r_{1,0}}= \left(0, \frac{(\gamma -\lambda_{\rm{R}}) (\gamma +\delta -1-\lambda_{\rm{R}})}{(\delta-1) (\gamma+\delta -1 -\delta  \lambda_{\rm{R}})}, 0, \frac{\gamma \lambda_{\rm{R}} (\gamma -\lambda_{\rm{R}}) (\gamma +\delta-1 -\lambda_{\rm{R}})}{(\delta -1) \delta  (\gamma+\delta -1 -\delta  \lambda_{\rm{R}})} \right), \nonumber \\   
&\ket{r_{1,1}}=\left(0, \frac{(1-\gamma ) (\gamma +\delta-1 -\lambda_{\rm{R}})}{\delta  (\gamma+\delta -1 -\delta \lambda_{\rm{R}})}, 0 , -\frac{(\gamma -1) \lambda_{\rm{R}} (\gamma-\lambda_{\rm{R}}) (\gamma +\delta-1 -\lambda_{\rm{R}})}{(\delta -1) \delta (\gamma+\delta -1 -\delta  \lambda_{\rm{R}})}  \right), \nonumber \\
&\ket{r'_0}=\left(0, \frac{\gamma  \lambda_{\rm{R}} (\gamma +\delta-1 -\lambda_{\rm{R}})}{\delta (\gamma +\delta -1-\delta  \lambda_{\rm{R}})}, 1, \frac{\lambda_{\rm{R}}^2(\gamma -\lambda_{\rm{R}})}{\gamma+\delta -1 -\delta  \lambda_{\rm{R}}}\right), \nonumber \\
&\ket{r'_1}=\left(0, \frac{(1-\gamma ) \lambda_{\rm{R}} (\gamma +\delta-1 -\lambda_{\rm{R}})}{\delta (\gamma +\delta -1-\delta  \lambda_{\rm{R}})},\frac{1}{\delta }-1, \frac{(1-\delta ) \lambda_{\rm{R}}^2 (\gamma -\lambda_{\rm{R}})}{\delta  (\gamma+\delta -1-\delta  \lambda_{\rm{R}})} \right).
\end{align}
Now in order to get a consistent solution on both the left and right boundary we demand the two pairs of the spectral parameters $\xi$ and $\omega$ in Eqs.~(\ref{nessspc1},\ref{nessspc2},\ref{nessspc3},\ref{nessspc4}) to be equal. This gives us a closed pair of equations for $\lambda_{\rm L}$ and $\lambda_{\rm R}$,
\begin{align}
&\frac{(\alpha +\beta -1)-\beta  \lambda_{\rm{L}}}{(\beta -1) \lambda_{L}^{2}}=\frac{\lambda_{\rm{R}} (\gamma -\lambda_{\rm{R}})}{(\delta -1) }, \\
&\frac{\lambda_{\rm{L}} (\alpha -\lambda_{\rm{L}})}{(\beta -1)}=\frac{ (\gamma +\delta -1)-\delta  \lambda_{\rm{R}}}{(\delta -1) \lambda_{\rm{R}}^2}.
\end{align}
Rewriting these equations in terms of the eigenvalue $\lambda=\lambda_{\rm{L}} \lambda_{\rm{R}}$ leads to
\begin{align}
&(\lambda-1)\times \label{char1}\\
&\left(\lambda ^3+\lambda ^2 (1-\alpha  \gamma )+\lambda  [\beta  \delta -(\alpha +\beta -1) (\gamma +\delta -1)]-(\alpha +\beta -1) (\gamma +\delta -1)\right)=0. \nonumber
\end{align}
Clearly, $\lambda=1$ is always the solution, corresponding to NESS. The remainder is a cubic polynomial. Thus there are also three other solutions corresponding to three decay modes whose eigenvalues do not change with system size. This set of four eigenvalues shall be referred to as the NESS-orbital.

Following the same procedure for the case of Bernoulli driving \eqref{bern} we find for the left boundary equations
\begin{align}
 &\xi=\frac{\lambda_{\rm{L}}+\mu \left(1-\lambda_{\rm{L}}\right)}{(\mu +1) \lambda_{\rm{L}}^2}, \\
 &\omega=\frac{\lambda_{\rm{L}} \left(2 \lambda_{\rm{L}}-1\right)}{ \mu +1},
 \end{align}
 and on the right,
 \begin{align}
 &\xi=\frac{\lambda_{\rm{R}} \left(2 \lambda _{\rm{R}}-1 \right)}{\sigma +1},\\
 &\omega=\frac{\lambda_{\rm{R}}+\sigma \left(1-\lambda_{\rm{R}}\right)}{\sigma \lambda_{\rm{R}}^2},
 \end{align}
 yielding the characteristic polynomial for the eigenvalue
\be
(\lambda -1)\left(4\lambda ^3+5 \lambda ^2- (\mu +\sigma +3)\lambda-\mu \sigma +2\right) =0. \label{char2}
\ee
The corresponding left boundary vectors are
\begin{align}
&\bra{l_0}=\scalemath{1}{\left( 0,\frac{2 (\lambda_{\rm{L}} (\mu -1)-\mu ) (\mu +\nu )}{\lambda_{\rm{L}}
   (2 \lambda_{\rm{L}}+\mu )}, \frac{\mu-\lambda_{\rm{L}}(\mu-1)}{\lambda_{\rm{L}}^2 (2 \lambda_{\rm{L}}-1)}, \frac{
   2 \lambda_{\rm{L}} (\nu +\mu  (\mu +\nu )+1)-(\mu -1) \mu}{2 \lambda_{\rm{L}}+\mu }  \right)}, \nonumber \\
&\bra{l_1}= \scalemath{1}{\left(0,\frac{2(\lambda_{\rm{L}} (\mu -1)-\mu ) (\mu +\nu +1)}{\lambda_{\rm{L}}(2 \lambda_{\rm{L}}+\mu )}, -\frac{\mu-\lambda_{\rm{L}}(\mu-1)}{\lambda_{\rm{L}}^2 (2 \lambda_{\rm{L}}-1)}, \frac{(\mu +1) [(2 \lambda_{\rm{L}}-1) \mu +2 \lambda_{\rm{L}} \nu ]}{2 \lambda_{\rm{L}}+\mu } \right)}, \nonumber\\  
&\bra{l'_{0,0}}= \left( \mu -\frac{\mu }{\lambda_{\rm{L}}}-1,
\mu ^2-2 (\lambda_{\rm L} (\mu-1) +1) (\mu +\nu )-1, 0, 0 \right), \nonumber \\
&\bra{l'_{0,1}}= \left(0, 0, \lambda_{\rm{L}} (2 \lambda_{\rm{L}}-1) (\mu -1), \mu-\frac{\mu}{\lambda_{\rm{L}}} -1\right), \nonumber \\
&\bra{l'_{1,0}}=\left( 2 \lambda_{\rm{L}} (\mu -1)+\frac{\mu }{\lambda_{\rm{L}}}-3 \mu +1, 2 \lambda_{\rm{L}} (\mu -1) (\mu +\nu +1)-\mu ^2+2 \nu +1, 0, 0 \right), \nonumber \\
&\bra{l'_{1,1}} =\left( 0, 0, -\lambda_{\rm{L}} (\mu +1), \mu-\frac{\mu }{\lambda_{\rm{L}}} -1  \right),
\end{align}
and the right boundary vectors are
\begin{align}
&\ket{r_{0,0}}= \left(0, -\lambda_{\rm{R}} (\sigma +1), -2 \rho  (\sigma  \lambda_{\rm{R}}-\lambda_{\rm{R}}+1)-\sigma  (2 \sigma  \lambda_{\rm{R}}-2 \lambda_{\rm{R}}-\sigma +2)-1 , 0 \right), \nonumber\\
&\ket{r_{0,1}}=\left(2 \lambda_{\rm{R}}^2 (\sigma +1), \lambda_{\rm{R}} (\sigma +1), -\sigma ^2+2\rho +2 \lambda_{\rm{R}} \left(\sigma ^2+\rho  \sigma -\rho -1\right)+1, 0  \right), \nonumber \\
&\ket{r_{1,0}}=\left(0, \lambda_{\rm{R}} (2 \lambda_{\rm{R}}-1) (\sigma -1), 0, -\lambda_{\rm{R}}^2 (2 \lambda_{\rm{R}}-1)\right), \nonumber \\ 
&\ket{r_{1,1}} =\left(0,-\lambda_{\rm{R}} (\sigma +1), 0, -\lambda_{\rm{R}}^2 (2 \lambda_{\rm{R}}-1) \right), \nonumber\\
&\ket{r'_0}=\left(-\lambda_{\rm{R}}^2,0, -\frac{\lambda_{\rm{R}} \left(2 \lambda_{\rm{R}} \left(\rho  \sigma +\rho +\sigma ^2+1\right)-\sigma ^2+\sigma \right)}{2 \lambda_{\rm{R}}+\sigma }, \frac{2 \lambda_{\rm{R}}^3 (2\lambda_{\rm{R}}-1) (\rho +\sigma )}{2 \lambda_{\rm{R}}+\sigma }\right), \nonumber\\
&\ket{r'_1}=\left( 0, -\lambda_{\rm{R}}^2, \frac{\lambda_{\rm{R}} (\sigma +1) (2\lambda_{\rm{R}} (\rho +\sigma )-\sigma )}{2 \lambda_{\rm{R}}+\sigma }, -\frac{2\lambda_{\rm{R}}^3 (2 \lambda_{\rm{R}}-1) (\rho +\sigma +1)}{2\lambda_{\rm{R}}+\sigma } \right).
\end{align}
In summary, the NESS and three other decay modes whose eigenvalue does not depend on the system size can be obtained from the compatibility condition of the `scattering' of MPA eigenvector (\ref{pness},\ref{ppness}) from the left and from the right stochastic boundary.

\section{Generalization of the bulk algebra and the decay modes}
\label{sec:firstorbital}
The bulk algebra (\ref{bulk1},\ref{bulk1b}) admits several generalizations, one of which allows us to construct a set of decay modes for the two types of stochastic boundary drivings. 
For this purpose we introduce an additional parameter $z\in\CC$, which will be referred to as {\em momentum parameter}, and define the following $4\times 4$ matrices
\begin{align}
&F_{+}=\left(
\begin{array}{cccc}
 0 & 0 & 0 & 0 \\
 0 & 0 & 0 & z \\
 0 & 0 & \frac{\xi  \omega -1}{\omega  z^2} & 0 \\
 0 & 0 & 0 & \xi  z^2 \\
\end{array} 
\right), \qquad\qquad\quad\; F_{-}=\left(
\begin{array}{cccc}
 0 & 0 & 0 & 0 \\
 0 & 0 & 0 & \frac{1}{\xi ^2 z^3} \\
 0 & 0 & \frac{\xi  \omega -1}{\xi  z^2} & 0 \\
 0 & 0 & 0 & \omega +\frac{1}{\xi}\left(\frac{1}{z^2}-1\right) \\
\end{array}
\right), \nonumber\\
&F'_{+}=\left(
\begin{array}{cccc}
 0 & 0 & 0 & 0 \\
 0 & 0 & 0 & \frac{z^3}{\omega ^2} \\
 0 & 0 & \frac{z^2 (\xi  \omega -1)}{\omega } & 0 \\
 0 & 0 & 0 & \xi +\frac{z^2-1}{\omega } \\
\end{array}
\right),\qquad F'_{-}= \left(
\begin{array}{cccc}
 0 & 0 & 0 & 0 \\
 0 & 0 & 0 & \frac{1}{z} \\
 0 & 0 & \frac{z^2 (\xi  \omega -1)}{\xi } & 0 \\
 0 & 0 & 0 & \frac{\omega }{z^2} \\
\end{array}
\right), \nonumber\\
&G_{+}=\left(
\begin{array}{cccc}
 0 & 0 & 0 & 0 \\
 0 & 0 & 0 & 0 \\
 \xi ^2 & 0 & 0 & 0 \\
 -\frac{2 \xi }{z} & -1 & 0 & 0 \\
\end{array}
\right), \qquad\qquad\quad\;\, G_{-}=\left(
\begin{array}{cccc}
 0 & 0 & 0 & 0 \\
 0 & 0 & 0 & 0 \\
 \xi  \omega  & 0 & 0 & 0 \\
 -\frac{\xi  \omega  z^2+1}{\xi  z^3} & -\frac{\omega }{\xi  z^2} & 0 & 0 \\
\end{array}
\right), \nonumber \\
&G'_{+}= \left(
\begin{array}{cccc}
 0 & 0 & 0 & 0 \\
 0 & 0 & 0 & 0 \\
 \xi  \omega  & 0 & 0 & 0 \\
 -z^3 \left(\frac{\xi }{z^2}+\frac{1}{\omega}\right) & -\frac{\xi  z^2}{\omega } & 0 & 0 \\
\end{array}
\right), \quad
G'_{-}=\left(
\begin{array}{cccc}
 0 & 0 & 0 & 0 \\
 0 & 0 & 0 & 0 \\
 \omega ^2 & 0 & 0 & 0 \\
 -2 \omega  z & -1 & 0 & 0 \\
\end{array}
\right),\nonumber \\
&K_{+}=\left(
\begin{array}{cccc}
 0 & 0 & 0 & 0 \\
 0 & \xi  z^2 & 0 & 0 \\
 0 & 0 & \frac{z^2 (\xi  \omega -1)}{\omega } & 0 \\
 0 & z & 0 & 0 \\
\end{array}
\right), \qquad\quad\;\;\; K_{-}=\left(
\begin{array}{cccc}
 0 & 0 & 0 & 0 \\
 0 & \omega +\frac{1}{\xi}\left(\frac{1}{z^2}-1\right) & 0 & 0 \\
 0 & 0 & \frac{z^2 (\xi  \omega -1)}{\xi } & 0 \\
 0 & \frac{1}{\xi ^2 z^3} & 0 & 0 \\
\end{array}
\right), \nonumber\\
&L_{+}=\left(
\begin{array}{cccc}
 0 & 0 & 0 & 0 \\
 0 & \xi  & 0 & -z \\
 \frac{\xi  \omega }{z} & 0 & 0 & 0 \\
 0 & 0 & -\frac{\xi  \omega +z^2}{\omega ^2 z} & -\frac{z^2}{\omega } \\
\end{array}
\right),\qquad\quad\;\, L_{-}=\left(
\begin{array}{cccc}
 0 & 0 & 0 & 0 \\
 0 & \frac{1}{\xi  z^2} & 0 & -\frac{\omega }{\xi  z} \\
 0 & 0 & \frac{\omega }{z^2} & 0 \\
 -2 \omega  & 0 & 0 & -\omega  \\
\end{array}
\right).
\end{align}
We will also need define objects which are vectors in physical space and matrices in a 8-dimensional auxiliary space. We define block diagonal operators (with physical space component $s=0,1$),
\begin{align}
&\hat{W}_s= e_{11}\otimes W_s(\xi z, \omega/z)+ e_{22}\otimes W_s(\xi/ z, \omega z), \nonumber \\
& \hat{W}'_s=e_{11}\otimes W'_s(\xi z, \omega/z)+ e_{22}\otimes W'_s(\xi/ z, \omega z), \label{diagpart}
\end{align}
where $e_{i j} = \ket{i}\bra{j}$, $i,j\in \{1,2\}$, are $2\times 2$ unit matrices.  
Further define the following block triangular $8\times 8$ matrices in extended auxiliary space
\begin{align}
&\hat{F}^{(k)}=\one_8+ e_{1 2} \otimes \frac{c_+ z^k F_+ +c_- z^{-k} F_-}{\xi \omega -1}, \quad \hat{F}'^{(k)}=\one_8+ e_{1 2} \otimes \frac{c_+ z^k F'_+ +c_- z^{-k} F'_-}{\xi \omega -1}, \nonumber \\
&\hat{G}^{(k)}=\one_8+ e_{1 2} \otimes \frac{c_+ z^k G_+ +c_- z^{-k} G_-}{\xi \omega -1}, \quad \hat{G}'^{(k)}=\one_8+ e_{1 2} \otimes \frac{c_+ z^k G'_+ +c_- z^{-k} G'_-}{\xi \omega -1}, \nonumber \\
&\hat{K}^{(k)}=\one_8+ e_{1 2} \otimes \frac{c_+ z^k K_+ +c_- z^{-k} K_-}{\xi \omega -1}, \nonumber\\
&\hat{L}^{(k)}=(z e_{11}+z^{-1} e_{22})\otimes \one_4+ e_{1 2} \otimes \frac{c_+ z^k L_+ +c_- z^{-k} L_-}{\xi \omega -1}. \label{FLhat} 
\end{align} 
which depend on a pair of complex amplitude parameters $c_+,c_-$.
We now state the generalized inhomogeneous bulk relations
\begin{align}
&P_{123}  \hat{K}^{(k-1)}\mathbf{\hat{W}}_1 \hat{S} \hat{G}^{(k)}\mathbf{\hat{W}}_2 \hat{F}^{(k+1)}\mathbf{\hat{W}}'_3=\hat{F}'^{(k-1)}\mathbf{\hat{W}}_1 \hat{G}'^{(k)} \mathbf{\hat{W}}'_2  \hat{K}^{(k+1)}\mathbf{\hat{W}}_3 \hat{S} \nonumber \\
&P_{123} \hat{G}'^{(k-1)}\mathbf{\hat{W}}'_1 \hat{F}'^{(k)} \mathbf{\hat{W}}_2  \hat{L}^{(k+1)}\mathbf{\hat{W}}'_3 \hat{S}=\hat{L}^{(k-1)}\mathbf{\hat{W}}'_1 \hat{S} \hat{F}^{(k)}\mathbf{\hat{W}}'_2 \hat{G}^{(k+1)}\mathbf{\hat{W}}_3, \label{bulk2}
\end{align}
where $\hat{S}=\one_2 \otimes S$, which can be straightforwardly checked to hold for any 
$\xi,\omega,z,c_+,c_-\in\CC$ and $k\in\ZZ$. Note that by setting $z=1, c_+=c_-=0$, we recover the original bulk algebra (\ref{bulk1},\ref{bulk1b}).
 
Defining a parity/swap transformation $\mathcal{R} : (\xi, \omega, z, c_+, c_-) \rightarrow (\omega, \xi, 1/z, c_-, c_+)$, $\mathcal{R}^2=\mathrm{id}$, we find
\be
\mathcal{R} \hat{F}^{(k)}=\hat{F}'^{(k)} \qquad \mathcal{R} \hat{G}^{(k)}=\hat{G}'^{(k)} \qquad \mathcal{R} \hat{\mathbf{W}}=\hat{\mathbf{W}}'.
\ee
Applying $\mathcal{R}$ to the bulk relations \eqref{bulk2} leads to another set of nonequivalent bulk relations (with   $\hat{K}'^{(k)}:= \mathcal{R} \hat{K}^{(k)}$ and $\hat{L}'^{(k)}:= \mathcal{R} \hat{L}^{(k)}$). 
There are numerous other similar extensions of the bulk algebra (\ref{bulk1},\ref{bulk1b}), but they do not seem to be useful for constructing eigenvectors for the boundary drivings studied, so we omit writing them here. 

\medskip
\noindent
{\bf Lemma:} Let us assume that 8-dimensional boundary vectors $\bra{\hat{l}_{s}}$, $\bra{\hat{l}'_{s,s'}}$, $\ket{\hat{r}_{s,s'}}$, $\ket{\hat{r}'_s}$ exist, together with parameters 
$\xi,\omega,z,c_+,c_-,\Lambda_{\rm L},\Lambda_{\rm R}$, such that the following boundary equations are satisfied 
\begin{align}
&P^{\rm{L}}_{12} \bra{\mathbf{\hat{l}}'_{12}}=\Lambda_{\rm{L}} \bra{\mathbf{\hat{l}}_1} \hmp{G}0 \hmup{W}{2}, \label{bound1} \\
&P_{123} \bra{\mathbf{\hat{l}}_{1}} \hm{L}{0}\hmup{W}{2}\hat{S} \hm{F}{1}\hmup{W}{3}=\bra{\mathbf{\hat{l}}'_{12}} \hm{K}{1}\hmu{W}{3}\hat{S},\label{bound2} \\
&P^{R}_{12} \ket{\mathbf{\hat{r}}_{12}}=\hm{F}{n-3}\hmup{W}{1}\hat{S}\ket{\mathbf{\hat{r}}'_2}, \label{bound3} \\
&P_{123} \hm{G}{n-4}\hmup{W}{1} \hm{K}{n-3}\hmu{W}{2}\hat{S} \ket{\mathbf{\hat{r}}'_{3}} =\Lambda_{\rm{R}} \hm{L}{n-4}\hmup{W}{1}\hat{S}\ket{\mathbf{\hat{r}}_{23}}. \label{bound4}
\end{align}
Then, the following inhomogeneous (site-dependent) MPA
\begin{align}
& \mathbf{p}=\bra{\mathbf{\hat{l}}_1}\hat{L}^{(0)}\WW'_2\hat{S}\hat{F}^{(1)}\mathbf{\WW}'_3\hat{G}^{(2)}\mathbf{\WW}_4 \cdots \hat{F}^{(n-5)}\mathbf{\WW}'_{n-3}\hat{G}^{(n-4)}\mathbf{\WW}_{n-2}\ket{\mathbf{\hat{r}}_{n-1,n}}, \label{p} \\
&  \mathbf{p'}=\bra{\mathbf{\hat{l}}'_{12}}\hat{F}^{'(1)}\mathbf{\WW}_3\hat{G}^{'(2)}\mathbf{\WW}'_4\hat{F}^{'(3)}\mathbf{\WW}_5 \cdots \hat{G}^{'(n-4)}\mathbf{\WW}'_{n-2}\hat{K}^{(n-3)}\mathbf{\WW}_{n-1}\hat{S}\ket{\mathbf{\hat{r}}'_{n}}, \label{pp}
\end{align}
generates an eigenvector of $U=U_{\rm e}U_{\rm o}$ (\ref{Ue},\ref{Uo}) with the eigenvalue $\Lambda=\Lambda_{\rm L}\Lambda_{\rm R}$.

\smallskip
\noindent
{\bf Proof:} 
Explaining how cancellation mechanism works is fully analogous to the simpler case of spatially homogeneous NESS-orbital (\ref{pness}-\ref{bound4a}). However, due to commutativity (\ref{boundcomm}) we can explain it here for the reverse order: When $U_{\rm e}$ acts on $\mathbf{p}$, $P_{123}$ first acts on the left boundary vector and via \eqref{bound2} creates $\hm{K}{1}\hmu{W}{3}\hat{S}$. The subsequent $P$'s in $U_{\rm e}$ transfer $\hm{K}{1}\hmu{W}{3}\hat{S}$ to the right via the bulk algebra \eqref{bulk2}. Before the final $P_{n-3,n-2,n-1}$ acts, $P_{\rm{R}}$ acts (as it commutes with $P_{n-3,n-2,n-1}$) and creates the $\hm{F}{n-3}\hmup{W}{1}\hat{S}$ necessary for the final $P_{n-3,n-2,n-1}$ to transfer $\hm{K}{n-5}\hmu{W}{n-3}\hat{S}$ to the end as $\hm{K}{n-3}\hmu{W}{n-1}\hat{S}$, thus finally creating $\mathbf{p'}$ \eqref{pp}. The odd-part of the propagator $U_{\rm o}$ acts analogously in reverse. 

\medskip
The MPA (\ref{p},\ref{pp}) will give us the leading decay mode and a set of other eigenvectors which we collectively call the \emph{first orbital}.
Due to the block upper triangular structure of $\hat{F}^{(k)},\hat{F}^{'(k)},\hat{G}^{(k)},\hat{G}^{'(k)}$, Eqs. (\ref{FLhat}), the matrix product state \eqref{p} (and analogously \eqref{pp}) can be written as a superposition of terms containing a string of $W_s(\xi z, \omega/z) W'_{s'}(\xi z, \omega/z)$ and then $c_\pm z^{\pm k} G_\pm$ (or $c_\pm z^{\pm k} F_\pm$) at just before $W$ (or $W'$) 
corresponding to site $k$ and then a string of $W_s(\xi /z, \omega z) W'_{s'}(\xi/ z, \omega z)$. There is also a boundary term from $L_\pm$ in the superposition. 
Therefore, the first orbital can be understood as single quasiparticle excitations over the NESS, which are composed as superpositions of left- and right-propagating waves $z^{\pm k}$ with 
non-trivial scattering at the boundaries. This is somewhat similar in form to the matrix coordinate ansatz used in solving the decay modes of the ASEP model \cite{matrixansatzASEP}. 

To solve the boundary equations we follow as similar procedure as outlined in Sec.~\ref{sec:NESS} for the NESS-orbital.
Let us decompose the extended auxiliary space as a direct sum of two 4-dimensional spaces ${\cal H}_1\oplus {\cal H}_2$, where
an element of ${\cal H}_i$ is written as $\ket{i}\otimes \ket{\psi}$.
Due to the upper triangular structure (\ref{FLhat}) the left boundary equations projected to the subspace ${\cal H}_1$ reduce to those for the NESS-orbital with a scaling factor $z$ coming from $e_{11}$ component of $\hat{L}^{(k)}$, but with rescaled spectral parameters ($\xi\to \xi z$, $\omega \to \omega/z$, because of \eqref{diagpart}). Since the NESS solution we found in Sec.~\ref{sec:NESS} is unique, the left boundary vector in this subspace must be the NESS-orbital boundary vector with scaled $\lambda_{\rm L} = \Lambda_{\rm L}/z$. Comparing  (\ref{bound1},\ref{bound2}) with (\ref{bound1a},\ref{bound2a}) and
(\ref{nessspc1},\ref{nessspc2}) immediately fixes the values of the spectral parameters $\xi$ and $\omega$ (writing first for the conditional driving (\ref{cond})):
\begin{align}
&\xi=\frac{z (\alpha +\beta -1)-\beta  \Lambda_{\rm{L}}}{(\beta -1) \Lambda_{L}^{2}}, \label{xi1}\\
&\omega =\frac{\Lambda_{\rm{L}} (\alpha  z-\Lambda_{\rm{L}})}{(\beta -1) z}. \label{paraleft1}
\end{align}
The remaining components of the left boundary equations in the subspace ${\cal H}_2$ come from either the diagonal components $e_{22}$ or the off-diagonal components $e_{12}$
of the auxiliary space operators. Requiring that the equations are solved for arbitrary $z, \alpha,\beta$, this fixes the ratio of the amplitudes
\begin{equation}
\frac{c_-}{c_+}=\frac{\Lambda_{L}^{4}}{z^4 (\alpha +\beta -1)}, \label{paraleft2}
\end{equation}
and that the general form of the left boundary vectors must be
\begin{align}
&\bra{\hat{l}_s}=\bra{1} \otimes \Big\langle l_s\Big(\lambda_{\rm{L}}=\frac{\Lambda_{\rm{L}}}{z}\Big)\Big| + c_+ \bra{2} \otimes \bra{\tilde{l}_s},\nonumber \\ 
&\bra{\hat{l}'_{s,s'}}=z \bra{1} \otimes \Big\langle l'_{s,s'}\Big(\lambda_{\rm{L}}=\frac{\Lambda_{\rm{L}}}{z}\Big)\Big|+ c_+ \bra{2} \otimes \bra{\tilde{l}'_{s,s'}}, \label{leftvecs}
\end{align}
where the explicit form of the `offdiagonal' vectors $\bra{\tilde{l}_s}$ and $\bra{\tilde{l}'_{s,s'}}$ are given in \ref{boundaryvecs}. Following a fully analogous procedure for the right boundary equations \eqref{bound3}, \eqref{bound4}, we arrive to the following results for the spectral parameters and the amplitude ratio
\begin{align}
&\xi=\frac{\Lambda_{\rm{R}} (\gamma  z-\Lambda_{\rm{R}})}{(\delta -1) z}, \label{xi2} \\
&\omega =\frac{z (\gamma +\delta -1)-\delta  \Lambda_{\rm{R}}}{(\delta -1) \Lambda_{\rm{R}}^2},\label{om2} \\
&\frac{c_+}{c_-}=\frac{\Lambda_{\rm{R}}^{4} z^{4 m+2}}{\gamma +\delta -1}, \label{pararight}
\end{align}
where $m=\frac{n}{2}-2$, and for the right boundary vectors
\begin{align}
&\ket{\hat{r}_{s,s'}}= c_- z^{n-3}\ket{1} \otimes \ket{\tilde{r}_{s,s'}} + \ket{2} \otimes \Big| r_{s,s'}\Big(\lambda_{\rm{R}}=\frac{\Lambda_{\rm{R}}}{z} \Big)\Big\rangle, \nonumber\\ 
&\ket{\hat{r}'_s}=c_- z^{n-3} \ket{1} \otimes \ket{\tilde{r}'_s} + \ket{2} \otimes \Big| r'_s\Big(\lambda_{\rm{R}}=\frac{\Lambda_{\rm{R}}}{z}\Big)\Big\rangle, \label{rightvecs}
\end{align}
whose components in ${\cal H}_2$ are expressed in terms of right boundary vectors for the NESS-orbital and the complementary (offdiagonal) components
$\ket{\tilde{r}_{s,s'}}$ and $\ket{\tilde{r}'_{s}}$ are given in \ref{boundaryvecs}.
Pairwise identifying equations (\ref{xi1},\ref{xi2}), (\ref{paraleft1},\ref{om2}), and (\ref{paraleft2},\ref{pararight}) we obtain a closed set of {\em Bethe-like} equations
for $\Lambda_{\rm L},\Lambda_{\rm R},z$:
\begin{align}
&\frac{z (\alpha +\beta -1)-\beta  \Lambda_{\rm{L}}}{(\beta -1) \Lambda_{L}^{2}}=\frac{\Lambda_{\rm{R}} (\gamma  z-\Lambda_{\rm{R}})}{(\delta -1) z}, \\
& \frac{\Lambda_{\rm{L}} (\alpha  z-\Lambda_{\rm{L}})}{(\beta -1) z}=\frac{z (\gamma +\delta -1)-\delta  \Lambda_{\rm{R}}}{(\delta -1) \Lambda_{\rm{R}}^2}, \\
&(\alpha +\beta -1)(\gamma +\delta -1)=\Lambda_{\rm{L}}^4 \Lambda_{\rm{R}}^4 z^{4m-2}.
\end{align}
Writing $\Lambda_{\rm L} = \Lambda/\Lambda_{\rm R}$ and
eliminating the variable $\Lambda_{\rm R}$ these equations can be rewritten as a pair of algebraic equations for $\Lambda,z$, the first of which can be understood as a 
nonequilibrium quasiparticle dispersion relation
\begin{align}
&\Lambda ^4-\alpha  \gamma  \Lambda ^3 z^2+ (\alpha +\beta +\gamma +\delta-\alpha\delta-\beta  \gamma-2)\Lambda ^2 z^2 \nonumber \\
&\quad -\beta  \delta  \Lambda  z^2+ (\alpha +\beta -1) (\gamma +\delta -1)z^4=0,\label{bethe1}\\
&(\alpha +\beta -1)(\gamma +\delta -1)-\Lambda^4 z^{4m-2}=0. \label{bethe2}
\end{align}
This equation has $4(2m+1)$ distinct roots which comprise the first orbital.

For the case of Bernoulli driving \eqref{bern} we find the following solutions for the left boundary equations (\ref{bound1},\ref{bound2}) 
\begin{align}
 &\xi=\frac{\Lambda_{\rm{L}}+\mu \left(z-\Lambda_{\rm{L}}\right)}{(\mu +1) \Lambda_{\rm{L}}^2}, \\
 &\omega=\frac{\Lambda_{\rm{L}} \left(2 \Lambda_{\rm{L}}-z\right)}{z (\mu +1)},\\
 &\frac{c_-}{c_+}=\frac{2 \Lambda_{\rm{L}}^4}{z^4 \mu},
 \end{align}
 and for the right ones (\ref{bound3},\ref{bound4})
 \begin{align}
 &\xi=\frac{\Lambda_{\rm{R}} \left(2 \Lambda _{\rm{R}}-z \right)}{z (\sigma+1)},\\
 &\omega=\frac{\Lambda_{\rm{R}}+\sigma \left(z-\Lambda_{\rm{R}}\right)}{(\sigma +1) \Lambda_{\rm{R}}^2},\\
 &\frac{c_+}{c_- }=\frac{2 z^{4m+2} \Lambda_{\rm{R}}^4}{\sigma }.
 \end{align}
Note that in this case these equations only depend on the difference of the driving parameters $\mu=\alpha-\beta$ (on the left) and $\sigma=\gamma-\delta$ (on the right) and thus so will the eigenvalues. 
 These lead to the following set of of Bethe-like equations,
  \begin{align}
&\frac{\Lambda_{\rm{L}}+\mu \left(z-\Lambda_{\rm{L}}\right)}{(\mu +1) \Lambda_{\rm{L}}^2}=\frac{\Lambda_{\rm{R}} \left(2 \Lambda_{\rm{R}} -z\right)}{z (\sigma +1)}\\
&\frac{\Lambda_{\rm{R}}+\sigma \left(z-\Lambda_{\rm{R}}\right)}{(\sigma +1) \Lambda_{\rm{R}}^2}=\frac{\Lambda_{\rm{L}} \left(2 \Lambda_{\rm{L}}-z\right)}{z (\mu +1)}\\
&\frac{1}{4} \mu \sigma= \Lambda_{\rm{L}}^4 \Lambda_{\rm{R}}^4 z^{4 m-2},
 \end{align}	
or equivalently
\begin{align}
&4 \Lambda ^4+\Lambda ^3 z^2+(\mu+\sigma -2)\Lambda ^2 z^2- (\mu -1) (\sigma -1)\Lambda  z^2+\mu \sigma z^4 =0 \label{bethe1a}, \\
& \frac{1}{4} \mu \sigma- \Lambda^4 z^{4 m-2}=0.\label{bethe2a}
\end{align}
The boundary vectors in the Bernoulli driving case are of the same form as in the conditional driving case, namely (\ref{leftvecs},\ref{rightvecs}). 
The explicit expressions for their components are given in \ref{boundaryvecs}. 

\subsection{Thermodynamics}
\label{subsec:thermo}
\begin{figure}
 \centering	
\vspace{-1mm}
\includegraphics[width=0.82\columnwidth]{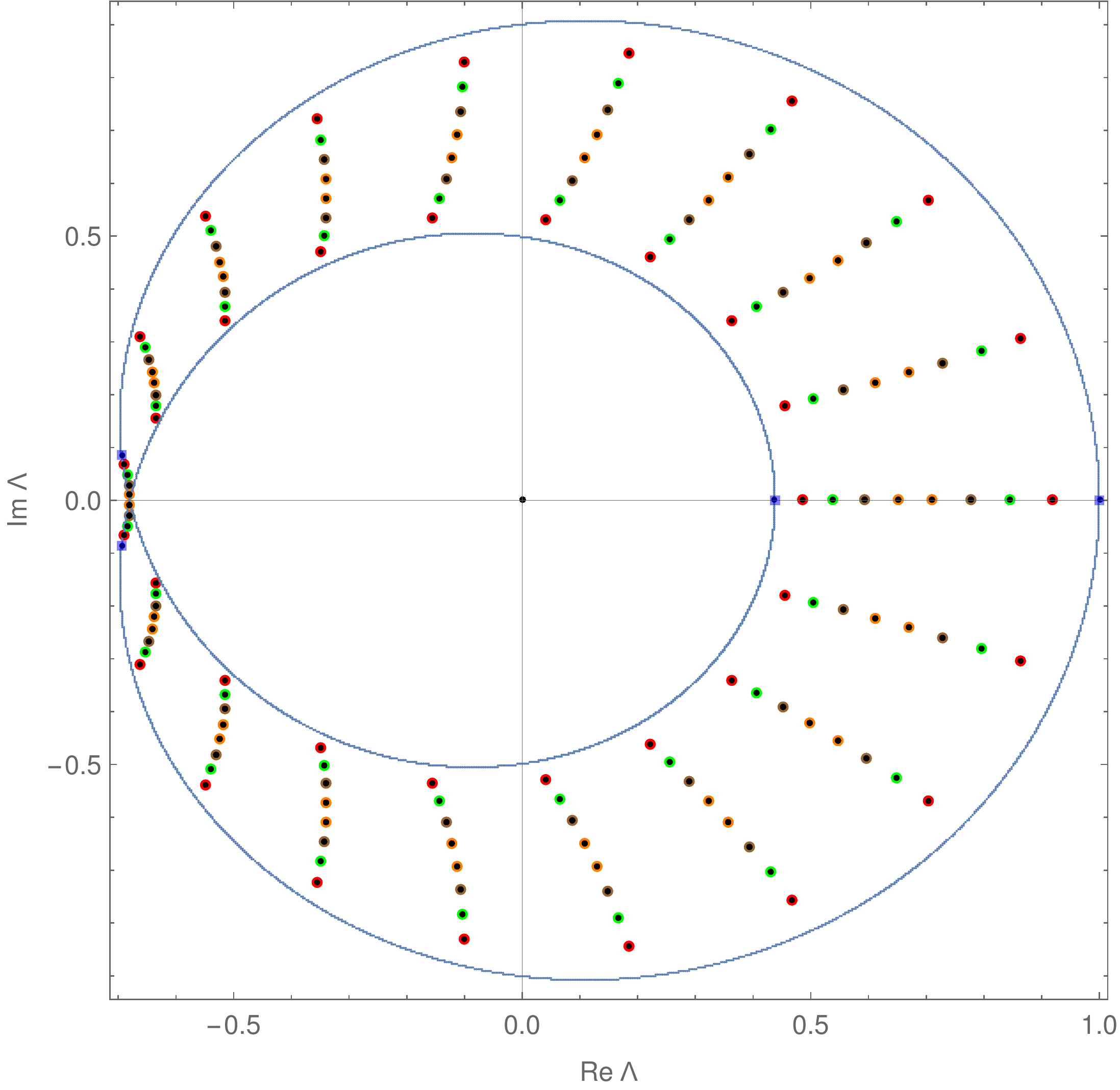}
\vspace{-1mm}
\caption{The Markov spectrum of the boundary driven Rule 54 cellular automaton for $n=12$ for conditional driving at $\alpha=1/4,\beta=1/3,\gamma=1/5,\delta=2/7$. The black dots show the numerical results. The red ($p=1$), green ($p=2$), brown ($p=3$), orange ($p=4$) points are solutions of the Bethe-like equations for the $p$-th orbital. The blue squares are the roots of characteristic polynomial for the NESS-orbital. The blue curve is the algebraic curve to which the first orbital converges in the thermodynamic limit.}
\label{fig:orbitals}
\end{figure}

In this subsection we will study the thermodynamic properties of the eigenvalues of the first orbital. Expanding equation \eqref{bethe2} (or \eqref{bethe2a}) in $1/m$ (recalling that $m=\frac{n}{2}-2$) gives us that in the leading order of $1/n$ ($(1/n)^0=1)$, $z$ must be unimodular
\be
z=e^{\ii \kappa}, \label{uniz}
\ee 
where $\kappa$ is the \emph{quasi-momentum} which may be restricted to interval $[0, \pi)$ as \eqref{bethe1} and \eqref{bethe2} (or \eqref{bethe1a} and \eqref{bethe2a}) are symmetric under the transformation $z \to -z$. Thus, in the leading order of $1/n$ expansion the spectrum in the first orbital converges to an algebraic curve $\Lambda(\kappa)$ given by \eqref{uniz} and \eqref{bethe1} (or \eqref{bethe1a}) (see Fig.~\ref{fig:orbitals})

The case $\Lambda(\kappa=0)=1$ corresponds to the eigenvalues which in the thermodynamic limit converge to $1$ and their scaling with $1/n$ will determine the asymptotic relaxation rate of the system. Writing the expansion of $\Lambda(0)$ as 
\be
\Lambda(0)=1-\Lambda_1(0)/n+\mathcal{O}(n^{-2})
\ee 
and inserting into \eqref{bethe1} (and \eqref{bethe1a}) we find that the gap closes as $1/n$ for both drivings. In the case of conditional driving (\ref{cond}) we find
\begin{equation}
\Lambda_1(0)=\frac{[1-(\alpha +\beta -1) (\gamma +\delta -1)] \log [(\alpha +\beta -1) (\gamma+\delta -1)]}{2 (\alpha +\beta -1) (\gamma +\delta -1)+\alpha  \gamma -\beta  \delta-2}, \label{ldmth}
\end{equation}
and in the case of Bernoulli driving (\ref{bern}) we find
\begin{equation}
\Lambda_1(0)=\frac{[4-\mu \sigma] \log \left[\frac{1}{4} \mu \sigma \right]}{(\mu +1) (\sigma +1)-9}, \label{ldmtha}
\end{equation}
where $\mu=\alpha-\beta$ and $\sigma=\gamma-\delta$.
This $1/n$ scaling of the gap is consistent with the results of Ref.~\cite{drivenbobenko} and ballistic transport.

\subsection{Quadratic form of the bulk algebra}

Even though the most elementary, partonic blocks of our exact solutions satisfy {\em cubic} algebraic relations, either (\ref{bulk1},\ref{bulk1b}) or (\ref{bulk2}), it is possible to rewrite them in terms of a quadratic algebra at the expense of defining 
auxiliary space operators which depend on two adjacent physical sites rather than one.

Let us define the following inhomogeneous 4-component vectors of $8\times 8$ matrices
\begin{align}
&\hat{\mathbf{Z}}_{12}^{(k)}=\hat{F}^{(2k-1)} \hat{\mathbf{W}}'_1 \hat{G}^{(2k)} \hat{\mathbf{W}}_2, \nonumber \\
&\hat{\mathbf{Z}}_{12}^{'(k)}=\hat{F}^{'(2k-1)} \hat{\mathbf{W}}_1 \hat{G}^{'(2k)} \hat{\mathbf{W}}'_2, \nonumber \\
&\hat{\mathbf{Y}}_{12}^{(k)}=\hat{K}^{(2k-1)} \hat{\mathbf{W}}_1 \hat{S} \hat{G}^{(2k)} \hat{\mathbf{W}}_2, \nonumber \\
&\hat{\mathbf{Y}}_{12}^{'(k)}=\hat{F}^{'(2k-1)} \hat{\mathbf{W}}_1 \hat{L}^{(2k)} \hat{\mathbf{W}}'_2 \hat{S}.
\end{align}
The cubic bulk algebra \eqref{bulk2} is then equivalent to the following quadratic algebra (formulated as 16-component vectors over four consecutive physical sites $1234$)
\begin{align}
&P_{123}\hat{\mathbf{Y}}_{12}^{(k)} \hat{\mathbf{Z}}_{34}^{(k+1)}=\hat{\mathbf{Z}}_{12}^{'(k)}\hat{\mathbf{Y}}_{34}^{(k+1)}, \nonumber \\
&P_{234} \hat{\mathbf{Z}}_{12}^{'(k)}\hat{\mathbf{Y}}_{34}^{'(k+1)}=\hat{\mathbf{Y}}_{12}^{'(k)} \hat{\mathbf{Z}}_{34}^{(k+1)}.
\end{align}
This quadratic formulation has perhaps some appeal as it is reminiscent of ZF algebra.\footnote{Essential differences to any meaningful formulation of ZF algebra still remain, most notably, our scattering operator $P$ has no `momentum' dependence. Perhaps this can be mended by some stochastic deformation of the model.}
The eigenvector MPA (\ref{p},\ref{pp}) now transforms to:
\begin{align}
& \mathbf{p}=\bra{\mathbf{\hat{l}}_{12}}\hat{\mathbf{Z}}_{34}^{(1)}\hat{\mathbf{Z}}_{56}^{(2)} \ldots \hat{\mathbf{Z}}_{n-3,n-2}^{(m)} \ket{\mathbf{\hat{r}}_{n-1,n}}, \\
&  \mathbf{p'}=\bra{\mathbf{\hat{l}}'_{12}}\hat{\mathbf{Z}}_{34}^{'(1)}\hat{\mathbf{Z}}_{56}^{'(2)} \ldots \hat{\mathbf{Z}}_{n-3,n-2}^{'(m)}\ket{\mathbf{\hat{r}}'_{n-1,n}},
\end{align}
where, as always $m=n/2-2$ and,
\begin{align}
&\bra{\mathbf{\hat{l}}_{12}}= \bra{\mathbf{\hat{l}}_{1}} \hat{L}^{(0)} \hat{\mathbf{W}}'_2 \hat{S}, \\
&\ket{\mathbf{\hat{r}}'_{n-1,n}}=\hat{K}^{(n-3)} \hat{\mathbf{W}}_{n-1} \hat{S} \ket{\mathbf{\hat{r}}'_{n}}.
\end{align}
Note that one can formulate and solve the boundary equations solely in terms of such two-site boundary vectors arriving to equivalent Bethe-like equations as in the previous section.

\section{Conjectures about the complete spectrum: modified Bethe-like equations and the degeneracy}
\label{sec:conjectures}

Despite numerous attempts we were not able to construct any other eigenvectors of $U$ beyond NESS-orbital and the first orbital.
Still, numerical inspections of the spectrum (see Fig.~\ref{fig:orbitals}) suggest that the structure of higher orbitals should be very similar, but the eigenvalues become degenerate with the degeneracy which quickly increases with the level of the orbital.
These comprise in total $2^{n-2}$ nonvanishing eigenvalues while the eigenvalue 0 is $3\times 2^{n-2}$ fold degenerate.

We were able to guess the Bethe-like equations which reproduce the entire spectrum. Introducing an integer $p$ which counts the orbital level and runs from $1$ to $m=n/2-2$, we postulate the modified Bethe-like equations,
either for the conditional driving (\ref{cond}):
\begin{align}
&\frac{z (\alpha +\beta -1)-\beta  \Lambda_{\rm{L}}}{(\beta -1) \Lambda_{L}^{2}}=\frac{\Lambda_{\rm{R}} (\gamma  z-\Lambda_{\rm{R}})}{(\delta -1) z}, \\
& \frac{\Lambda_{\rm{L}} (\alpha  z-\Lambda_{\rm{L}})}{(\beta -1) z}=\frac{z (\gamma +\delta -1)-\delta  \Lambda_{\rm{R}}}{(\delta -1) \Lambda_{\rm{R}}^2}, \\
&(\alpha +\beta -1)^p(\gamma +\delta -1)^p=(\Lambda_{\rm{L}} \Lambda_{\rm{R}})^{4 p} z^{4(m-p)+2},
\end{align}
or for the Bernouli driving (\ref{bern}):
\begin{align}
&\frac{\Lambda_{\rm{L}}+\mu \left(z-\Lambda_{\rm{L}}\right)}{(\alpha -\beta +1) \Lambda_{\rm{L}}^2}=\frac{\Lambda_{\rm{R}} \left(2 \Lambda_{\rm{R}} -z\right)}{z (\sigma +1)},\\
&\frac{\Lambda_{\rm{R}}+\sigma \left(z-\Lambda_{\rm{R}}\right)}{(\sigma +1) \Lambda_{\rm{R}}^2}=\frac{\Lambda_{\rm{L}} \left(2 \Lambda_{\rm{L}}-z\right)}{z (\mu +1)},\\
&4^{-p} \mu^p \sigma^p=(\Lambda_{\rm{L}} \Lambda_{\rm{R}})^{4 p} z^{4(m-p)+2}.
\end{align}
Each of these orbitals has exactly $4(2m+1)$ distinct roots. We conjecture that the above equations, together with the NESS characteristic polynomial (\ref{char1},\ref{char2}), describe the entire spectrum of $U$.
Indeed, agreement with the spectrum obtained from numerical diagonalization of $U$, for up to $n=16$, is perfect, within trustable precision of numerical routines (as demonstrated in Fig.~\ref{fig:orbitals}).

Furthermore, we provide a conjecture for the average degeneracy of the eigenvalues of the $p$-th orbital which seems to scale as
\begin{equation}
g(m,p)=\frac{1}{p} \left(
\begin{array}{c} 
2 m \\ 
p-1 
\end{array}\right).
\end{equation}
Note that the function $g(m,p)$ can also be non-integer for some values of $m,p$. This means that there are different degeneracies within the same orbital $p$ for system size $n=2(m+2)$ and $g(m,p)$ gives the average value of the degeneracy. This conjecture has also been confirmed numerically for up to $n=16$.  We can make a simple consistency check by counting the total number of roots in all the orbitals together with their multiplicity $\sum_{p=1}^m 4 (2m+1) g(m,p) = 4 (4^m-1) = 2^{n-2}-4$ which together with four eigenvalues of the NESS-orbital yield the total number of $2^{n-2}$ nonvanishing eigenvalues.

\section{Conclusions and open problems}
\label{sec:conclusions}
We found that the spectrum of the Markov matrix of a deterministic boundary driven cellular automaton (the Rule 54) organizes into orbitals. Throughout the paper we gave results for two types of non-equivalent stochastic boundary drivings, \eqref{cond} and \eqref{bern}. 
The eigenvalues in each orbital fulfil a set of three coupled Bethe-like equations. We found explicit matrix product forms of the eigenvectors in two main orbitals -- the NESS-orbital (containing the NESS and three other eigenvectors) and the first orbital (which contains the leading decay mode and eigenvalues of the largest modulus). 
To find the NESS-orbital we used a 4-dimensional representation (with two spectral parameters) of a three-site bulk algebra cancellation mechanism \eqref{bulk1}.  This three-site bulk algebra is similar in form to a two-site bulk cancellation mechanism in discrete time ASEP models \cite{asep1, asep2, asep3}. 
To find the first orbital we generalized the aforementioned bulk algebra to an 8-dimensional auxiliary space \eqref{bulk2}. The structure of such positionally dependent bulk algebra allows for construction of the eigenvectors in a compact form of an inhomogeneous MPA, which may be understood as a superposition of local single-particle excitations over the NESS-orbital with different momenta, reminiscent of the matrix coordinate ansatz for ASEP models \cite{matrixansatzASEP}. We also investigated the thermodynamic properties and proved that the spectral gap yielding the ultimate relaxation time scales as $~1/n$. In the thermodynamic limit, the first orbital defines an algebraic curve in the complex plane which borders the spectrum of the model.
We have also shown how the cubic bulk algebra may be rewritten as a quadratic algebra with operators acting on two physical sites, instead of one. We note that our inhomogeneous MPA can be rewritten as well in the form of an inhomogeneous patch state ansatz as proposed and used to find NESS of the model in Ref.~\cite{drivenbobenko}, however the elements of the patch tensors have to be replaced by positionally dependent $2\times 2$ matrices. This has been actually the route through which we arrived at the results reported in the present paper.

Although we were unable to find all the eigenvectors, we provided a conjecture for the Bethe equations for all the orbitals, which has been corroborated by extensive numerical investigation (Sec.~\ref{sec:conjectures}). The higher orbitals are highly degenerate and we provide a conjecture for the average degeneracy of each orbital in the same section.

Many open questions remain. Most pressingly, one would wish to construct the eigenvectors in all the orbitals exactly and prove the conjecture for the general Bethe-like equations as given in Sec.~\ref{sec:conjectures}. A direct generalization of the bulk algebra \eqref{bulk2} to two-particle excitations does not seem to work, and numerous other generalizations are possible making it difficult to ascertain how to continue. The fact that the higher orbital Bethe-like equations are very similar to the first orbital's might suggest that the
nonequilibrium quasiparticle excitations are non-interacting and that $z$ represents simple the center-of-mass momentum of all excitations. However, such an idea seems to be an oversimplification since the independent particle model cannot explain exponentially large (in orbital level) degeneracy of the eigenvalues. We thus interpret this as an interacting model with an exponential {\em bunching} of quasiparticles. It is also not clear at present if and how the exact solution of the boundary driven Rule 54 model reported here  connects to Yang-Baxter equation and common language of integrability.

Another interesting question which remains is whether the cases (\ref{cond}) and (\ref{bern})
complete the set of integrable stochastic boundaries of the model or not. The problem of classification of integrable stochastic boundaries of a bulk deterministic (e.g. Hamiltonian) integrable theory is generally open.

Even though we were unable to construct the complete set of eigenvectors of our model we feel that the results obtained here can be extended for use in other models, for instance, to complement the study of asymptotic decay of densities in reaction models \cite{reacmodel}, help in the study of the decay modes of discrete time models, notably discrete time ASEP models \cite{asep1,asep2,asep3}, and in particular, to other driven integrable cellular automata with deterministic bulk dynamics (e.g., \cite{Bobenko,intCA1,intCA2,intCA3,intCA4}) and perhaps even driven quantum models in discrete time \cite{YF}. 

\section*{Acknowledgements}

We thank V. Popkov, M. Vanicat, E. Ragoucy for discussions and E. Ilievski and L. Zadnik for useful comments on the manuscript. The work has been supported by the grants P1-0044 and N1-0025 of Slovenian Research Agency, and ERC Grant OMNES.

\appendix 

\section{Boundary vectors}
\label{boundaryvecs}
In this appendix we list the so-called off-diagonal components of the boundary vectors which solve the boundary equations (\ref{bound1}-\ref{bound4}). We note that the values of all other parameters (which enter into the equations in a nonlinear way) and other (`diagonal') components of the boundary vectors can be fixed by comparing to a $4-$dimensional auxiliary problem for the NESS-orbital. The remaining components reported below are then fixed by solving the remaining system of {\em linear} equations. They are
quite lengthy and their algebraic form could probably still be considerably optimized, but they are given here just for completeness.

\subsection{Boundary vectors for conditional driving}
The offdiagonal left boundary vector components for the first orbital \eqref{leftvecs} and for conditional driving \eqref{cond} are given as
\bea
&\hspace{-25mm}\bra{\tilde{l}_{0}}=  C_1 \bigg(0,\frac{(\beta -1)^2 \beta  \Lambda_{\rm{L}} z \left(\Lambda_{\rm{L}}-z^3 (\alpha+\beta -1)\right)}{\left(z^2-1\right) (z (\alpha +\beta -1)-\Lambda_{\rm{L}})} , \\
&c_1 , -\frac{(\beta -1)^2 \beta  \Lambda_{\rm{L}}^2 z^2 \left(\beta  \Lambda_{\rm{L}}^2+\alpha ^2 z^2-\alpha  \Lambda_{\rm{L}} z \left((\beta -1) z^2+2\right)\right)}{\left(z^2-1\right) (\alpha  z-\Lambda_{\rm{L}}) (\alpha  z-\beta  \Lambda_{\rm{L}})} \bigg),\nonumber 
\eea
\bea
&\hspace{-25mm}\bra{\tilde{l}_{1}} = C_1\bigg(0 , \frac{(\beta -1)^3 \Lambda_{\rm{L}} z \left(z^3 (\alpha +\beta -1)-\Lambda_{\rm{L}}\right)}{\left(z^2-1\right) (z (\alpha +\beta -1)-\Lambda_{\rm{L}})} ,  \\
&c_2 , \frac{(\beta -1)^3 \Lambda_{\rm{L}}^2 z^2
   \left(\beta  \Lambda_{\rm{L}}^2+\alpha ^2 z^2-\alpha  \Lambda_{\rm{L}} z
   \left((\beta -1) z^2+2\right)\right)}{\left(z^2-1\right) (\alpha  z-\Lambda_{\rm{L}}) (\alpha  z-\beta  \Lambda_{\rm{L}})} 
    \bigg), \nonumber 
\eea
\bea
&\hspace{-25mm}\bra{\tilde{l}'_{0,0}} =\left( \frac{\alpha  \Lambda_{\rm{L}} (\Lambda_{\rm{L}}+z (1-\alpha -\beta))}{\beta  z
   \left(z^2-1\right) (\alpha +\beta -1)} , \frac{\Lambda_{\rm{L}}^2 (\Lambda_{\rm{L}}+z (1-\alpha -\beta))}{\left(z^2-1\right) (\alpha +\beta -1)} , 0 , 0 \right), 
\eea
\bea
&\hspace{-25mm}\bra{\tilde{l}'_{0,1}}=\left(0 , 0 , c_3, \frac{\alpha  \Lambda_{\rm{L}}(\Lambda_{\rm{L}}+z (1-\alpha -\beta))}{\beta  z \left(z^2-1\right) (\alpha+\beta -1)} \right),
\eea
\bea
&\hspace{-25mm}\bra{\tilde{l}'_{1,0}}= \left(\frac{\Lambda_{\rm{L}} (\alpha -\Lambda_{\rm{L}} z) (z (\alpha +\beta
   -1)-\Lambda_{\rm{L}})}{\beta  z \left(z^2-1\right) (\alpha +\beta -1)} ,
   \frac{(\beta -1) \Lambda_{\rm{L}}^2 (z (\alpha +\beta -1)-\Lambda_{\rm{L}})}{\beta  \left(z^2-1\right) (\alpha +\beta -1)} , 0 , 0 \right),
\eea
\bea
&\hspace{-25mm}\bra{\tilde{l}'_{1,1}} =\left(0 , 0 , c_4 , \frac{(\alpha -1)\Lambda_{\rm{L}} (z (\alpha +\beta -1)-\Lambda_{\rm{L}})}{\beta  z \left(z^2-1\right) (\alpha +\beta -1)}\right),
\eea
where
\be
C_1=\frac{1}{(\beta-1) (\alpha + \beta-1) \beta}, \nonumber
\ee
\bea
&\begin{split}
&c_1= \beta -\frac{\alpha ^2 (\beta -1) z^4}{(\Lambda_{\rm{L}}-\alpha  z)^2}+\frac{\alpha  z \left(\alpha +\alpha ^2 z^2-\alpha  z^2-\beta ^2 z^2\right)}{\beta  (\alpha +\beta ) (\beta  \Lambda_{\rm{L}}-\alpha  z)}-1\\
&+\frac{\alpha }{\beta(1 - z^2)}-\frac{z(\alpha +\beta -1)^2 \left(\alpha +\beta +z^2 (\alpha +\beta +1)-1\right)}{(\alpha+\beta ) (z (\alpha +\beta -1)-\Lambda_{\rm{L}})}\\
&+\frac{\alpha  z \left(-\alpha  \beta +\alpha +\beta +\beta  z^4 (\alpha +2 \beta-3)+z^2 \left(-\alpha -2 \beta ^2+\beta +2\right)-1\right)}{\beta  \left(z^2-1\right)(\alpha  z-\Lambda_{\rm{L}})},
\end{split} \nonumber
\eea
\bea
&\begin{split}
&c_2=\frac{\alpha ^2 (\beta -1)^2 z^4}{(\alpha  z-\Lambda_{\rm{L}})^2}+\frac{z \left(\alpha^2-\beta +z^2 \left(\alpha ^3-(\alpha +1) \beta ^2+\beta \right)\right)}{(\alpha+\beta ) (\alpha  z-\beta  \Lambda_{\rm{L}})}\\
&+\frac{(\beta -1) z (\alpha +\beta-1)^2 \left(\alpha +\beta +z^2 (\alpha +\beta +1)-1\right)}{(\alpha +\beta ) (z(\alpha +\beta -1)-\Lambda_{\rm{L}})}\\
&+\frac{z \left[(\alpha -1) (\alpha  (\beta -2)+1)-\alpha (\beta -1) z^4 (\alpha +2 \beta -3)\right]}{\left(z^2-1\right) (\alpha  z-\Lambda_{\rm{L}})}\\
&+\frac{z\left[z^2 (\alpha  (\alpha +\beta  (2 \beta-3)-1)-\beta +2)\right]}{\left(z^2-1\right) (\alpha  z-\Lambda_{\rm{L}})}-\frac{\alpha +(\beta -2) \beta +(\beta -1)^2 z^2}{z^2-1},
\end{split}\nonumber
\eea
\bea
\begin{split}
&c_3=-\frac{\Lambda_{\rm{L}}^3 (\alpha  z-\Lambda_{\rm{L}}) \left(\beta \Lambda_{\rm{L}}^2+z^4 (\alpha +\beta -1)^2-\Lambda_{\rm{L}} z (\alpha +\beta -1)\left(\beta +z^2\right)\right)}{z \left(z^2-1\right) (\alpha +\beta -1) (z (\alpha+\beta -1)-\beta  \Lambda_{\rm{L}})^2},
\end{split}\nonumber
\eea
\bea
\begin{split}
&c_4= \frac{\Lambda_{\rm{L}}^3(\alpha -1) (\beta -1)  (z (\alpha +\beta-1)-\Lambda_{\rm{L}}) }{\beta  \left(z^2-1\right) (\alpha +\beta -1) (\alpha  z-\Lambda_{\rm{L}}) (z (\alpha +\beta -1)-\beta  \Lambda_{\rm{L}})^2} \times \\
&\left[\alpha  z^4 (\alpha +\beta -1)-\beta  \Lambda_{\rm{L}}^2 \left(z^2-2\right)-\Lambda_{\rm{L}} z (\alpha  \beta +\alpha +\beta-1)\right].
\end{split}   \nonumber
\eea
The offdiagonal right boundary vectors components \eqref{rightvecs} are
\bea
&\hspace{-15mm}\ket{\tilde{r}_{0,0}}=C_2\left( \frac{\gamma  (1-\delta ) \Lambda_{\rm{R}}}{\delta  z \left(z^2-1\right) (\gamma +\delta -1) (z (\gamma +\delta -1)-\delta 
   \Lambda_{\rm{R}})} , 0 , c_5 , 0 \right), 
\eea
\bea
&\hspace{-15mm}\ket{\tilde{r}_{0,1}}=C_2 \left( \frac{(\delta -1) \Lambda_{\rm{R}} (\gamma -\Lambda_{\rm{R}} z) }{\delta  z \left(z^2-1\right) (\gamma +\delta -1) (z (\gamma +\delta -1)-\delta  \Lambda_{\rm{R}})} , 0 , c_6 , 0\right), 
\eea
\bea
&\hspace{-15mm}\ket{\tilde{r}_{1,0}}= C_2 \bigg( 0 , \frac{\Lambda_{\rm{R}} (\Lambda_{\rm{R}} z-\gamma )}{z \left(z^2-1\right) (\gamma +\delta -1) (z (\gamma +\delta
   -1)-\delta  \Lambda_{\rm{R}})} , 0 , \\
&\frac{\gamma  \Lambda_{\rm{R}}^3(\Lambda_{\rm{R}} z-\gamma ) (z (\gamma +\delta -1)-\Lambda_{\rm{R}})}{\delta \left(z^2-1\right) (\gamma +\delta -1) (z (\gamma +\delta -1)-\delta  \Lambda_{\rm{R}})} \bigg),\nonumber 
\eea
\bea
&\hspace{-15mm}\ket{\tilde{r}_{1,1}}= C_2\bigg(0 , \frac{(\gamma -1) (\delta -1) \Lambda_{\rm{R}}}{\delta  z \left(z^2-1\right) (\gamma +\delta -1) (z (\gamma
   +\delta -1)-\delta  \Lambda_{\rm{R}})} , 0 , \\
&\frac{(\gamma -1) \Lambda_{\rm{R}}^3(\gamma -\Lambda_{\rm{R}} z) (z (\gamma +\delta -1)-\Lambda_{\rm{R}})}{\delta \left(z^2-1\right) (\gamma +\delta -1) (z (\gamma +\delta -1)-\delta  \Lambda_{\rm{R}})} \bigg),\nonumber
\eea
\bea
&\hspace{-15mm}\ket{\tilde{r}_{0}}= C_3 \left( 0 , c_7 , (\delta -1) \delta \Lambda_{\rm{R}}^2 \theta , (\delta -1) \delta  \Lambda_{\rm{R}}^4 z^2 (\Lambda_{\rm{R}}-\gamma  z)^2 (\delta \Lambda_{\rm{R}} z-\gamma) \right), 
\eea
\bea
&\hspace{-15mm}\ket{\tilde{r}_{1}} =C_3 \left( 0 , \frac{1-\gamma}{\gamma}  c_7 ,-(\delta -1)^2 \Lambda_{\rm{R}}^2 \theta , (\delta -1)^2 \Lambda_{\rm{R}}^4 z^2 (\Lambda_{\rm{R}}-\gamma  z)^2 (\gamma-\delta  \Lambda_{\rm{R}} z) \right),
\eea
where,
\bea
&C_2=\Lambda_{\rm{R}} (z (\gamma +\delta-1)-\Lambda_{\rm{R}}), \nonumber\\
&C_3=\delta  z^2 \left(z^2-1\right) (\gamma +\delta -1) (\gamma  z-\Lambda_{\rm{R}})
   (\gamma  z-\delta  \Lambda_{\rm{R}}) (z (\gamma +\delta -1)-\delta  \Lambda_{\rm{R}}), \nonumber
\eea
\bea
&\hspace{-25mm}c_5=\frac{-\delta  \Lambda_{\rm{R}}^2 z^3+\Lambda_{\rm{R}}
   \left[\gamma +\delta +z^4 (\gamma +\delta -1)+(\gamma -1) (\delta -1)
   z^2-1\right)]+\gamma  z (-\gamma -\delta +1)}{z^2 \left(z^2-1\right) (\gamma
   +\delta -1) (\gamma  z-\Lambda_{\rm{R}}) (z (\gamma +\delta -1)-\delta 
   \Lambda_{\rm{R}})}, \nonumber \\
&\hspace{-25mm}c_6= 
\frac{(1-\delta ) \left(\Lambda_{\rm{R}} \left[\gamma +\delta +z^4 (\gamma\!+\!\delta\!-\!1)+(\gamma -1) (\delta -1) z^2-1\right]-\delta \Lambda_{\rm{R}}^2 z^3-\gamma  z (\gamma\!+\!\delta\!-\!1)\right)}{\delta
    z^2 \left(z^2-1\right) (\gamma +\delta -1) (\gamma  z-\Lambda_{\rm{R}}) (z (\gamma
   +\delta -1)-\delta  \Lambda_{\rm{R}})}, \nonumber \\
&\hspace{-25mm}c_7=\gamma  (1-\delta ) \Lambda_{\rm{R}}^3 z^2 (\gamma  z-\Lambda_{\rm{R}}) (z (\gamma
   +\delta -1)-\Lambda_{\rm{R}}) (\gamma  z-\delta  \Lambda_{\rm{R}}), \nonumber
\eea
\begin{eqnarray*}
&\hspace{-15mm}\theta= -\gamma ^2 z^6 (\gamma +\delta -1)+\delta  \Lambda_{\rm{R}}^3 z \left((\delta -1) z^2+1\right)\\
   &\hspace{-15mm}+\Lambda_{\rm{R}}^2 \left((\delta -1)^2+z^4
   (-(\gamma  (\delta -1) \delta +\gamma +\delta -1))+z^2 \left(-3 \gamma  \delta
   +\gamma -2 (\delta -1)^2\right)\right) \\
  &\hspace{-15mm}+\gamma  \Lambda_{\rm{R}} z^3 \left(\gamma 
   (2 \delta -1)+(\delta -1)^2+2 z^2 (\gamma +\delta -1)\right).
\end{eqnarray*}

\subsection{Boundary vectors for Bernoulli driving}

For Bernoulli driving \eqref{bern} the offdiagonal left boundary vector components of the first orbital \eqref{leftvecs} are
\bea
&\hspace{-25mm}\bra{\tilde{l}_{0}} =C_1 \left(
 0 , c_1 , c_2 , \chi_0 \frac{z^2 (z-2\Lambda_{\rm{L}})\Lambda_{\rm{L}}^2 (2
  \Lambda_{\rm{L}}+z \mu ) }{-\mu \Lambda_{\rm{L}}+\Lambda_{\rm{L}}+z \mu }\right), 
\eea
\bea
&\hspace{-25mm}\bra{\tilde{l}_{1}}=C_1 \left(  0 , -c_1 \left(1+\frac{1}{\mu +\nu }\right) , c_3 , \chi_1 \frac{z^2 (z-2
  \Lambda_{\rm{L}})\Lambda_{\rm{L}}^2 (\mu +1) (2\Lambda_{\rm{L}}+z \mu )}{-\mu 
  \Lambda_{\rm{L}}+\Lambda_{\rm{L}}+z \mu }\right),
\eea
\bea
&\hspace{-25mm}\bra{\tilde{l}'_{0,0}}=C_2\left(\frac{(-\mu \Lambda_{\rm{L}}+\Lambda_{\rm{L}}+z \mu )^2}{\Lambda_{\rm{L}}} ,\Lambda_{\rm{L}} (\mu -1) \kappa , 0 , 0 \right),
\eea
\bea
&\hspace{-25mm}\bra{\tilde{l}'_{0,1}}= C_2 \left(  0 , 0 , (z-2\Lambda_{\rm{L}})\Lambda_{\rm{L}} (\mu -1) \left(\mu 
   z^3+\Lambda_{\rm{L}}-\Lambda_{\rm{L}} \mu \right) , \frac{(-\mu  \Lambda_{\rm{L}}+\Lambda_{\rm{L}}+z \mu )^2}{\Lambda_{\rm{L}}} \right),
\eea
\bea
&\hspace{-25mm}\bra{\tilde{l}'_{1,0}}=C_2 \left( \frac{(2 z\Lambda_{\rm{L}}-1) (-\mu \Lambda_{\rm{L}}+\Lambda_{\rm{L}}+z \mu
   )^2}{\Lambda_{\rm{L}}} , c_5 , 0 , 0 \right),
\eea
\bea
&\hspace{-25mm} \bra{\tilde{l}'_{1,1}}=C_2 \bigg(  0, 0 , \frac{z\Lambda_L (\mu +1) \left[\mu  z^4-2\Lambda_L^2 (\mu
   -1) z^2+(\Lambda_L-3\Lambda_L \mu ) z+4\Lambda_L^2 (\mu
   -1)\right]}{z-2\Lambda_L} , \\
   &\frac{(-\mu \Lambda_L+\Lambda_L+z \mu )^2}{\Lambda_L}\bigg), \nonumber
\eea
where  
\bea
&C_1=\frac{(\mu +1)^2}{\mu  z^2 \left(z^2-1\right) (z-2\Lambda_{\rm{L}})^2 (\Lambda_{\rm{L}} (\mu -1)+z) (2\Lambda_{\rm{L}}+\mu z)^2 }, \nonumber \\
&C_2=\frac{\Lambda_{\rm{L}}^2 (\mu +1) [\Lambda_{\rm{L}} (\mu -1)-\mu ] (2
  \Lambda_{\rm{L}}+\mu  z -\Lambda_{\rm{L}} \mu)}{\mu  z^2 \left(z^2-1\right) (2\Lambda_{\rm{L}}+\mu ) \nonumber
   (\Lambda_{\rm{L}}+\mu  z)^2},
\eea
\be
\hspace{-13mm}c_1=2 z -\Lambda_{\rm{L}} (\mu +\nu ) (z-2\Lambda_{\rm{L}})^2\left(\mu  z^3+2 \Lambda_{\rm{L}}\right) (z \mu -\mu \Lambda_{\rm{L}}+\Lambda_{\rm{L}}) (z-\Lambda_{\rm{L}}+\Lambda_{\rm{L}} \mu ), \nonumber
\ee
\begin{align*}
&c_2= -\mu ^3 z^9+\Lambda_{\rm{L}} \mu ^2 \left(-2 \mu ^2+5 \mu -1\right) z^8+2\Lambda_{\rm{L}}^2 \mu  \left[6 \mu ^3-2 \mu^2+5\mu+2 (\mu +1)^2 \nu +1\right] z^7 \\
	&+\Lambda_{\rm{L}} \mu  \left[2 \left(\mu
   ^4-10 \mu ^3-6 \mu ^2-16 \mu -8 (\mu +1)^2 \nu -5\right)\Lambda_{\rm{L}}^2+(\mu-4) \mu  (\mu +1)\right] z^6 \\
   &-4\Lambda_{\rm{L}}^4\left[\mu ^5-\mu ^4-2 \mu
   ^2+\mu +\left(1-\mu ^2\right)^2 \nu +1\right] z^5\\
   &-\Lambda_{\rm{L}}^2\mu  (\mu +1) [4
   \nu +\mu  (11 \mu +4 \nu -13)+10] z^5\\
   &+2\Lambda_{\rm{L}} \left(\mu
   ^3-\Lambda_{\rm{L}}^2 (\mu +1) \left[\mu  \left\{\mu  \left(\mu ^2-15 \mu -10 \nu
   +5\right)-10 \nu -19\right\}+2\right]\right) z^4 \\
  &+2\Lambda_{\rm{L}}^2 \left( \mu ^4-4\mu ^3+3 \mu ^2 \right)z^3\\
  &+4\Lambda_{\rm{L}}^4 (\mu +1) \left\{2 \nu +\mu  (4-3 \nu +\mu  [4-4 \nu+\mu  (\mu +\nu -6)]-7)\right\} z^3 \\
   &+2\Lambda_{\rm{L}}^3 \left[4 \left(\mu
   ^2-1\right) \left(\mu ^2+\nu  \mu +\nu +1\right)\Lambda_{\rm{L}}^2+\mu  \left(2-5
   \mu ^3+4 \mu ^2-13 \mu -2 (\mu +1)^2 \nu\right)\right] z^2 \\
   &+4\Lambda_{\rm{L}}^4 \left[(3 \mu -1) \nu  (\mu +1)^2+4 \mu  \left(\mu ^3+2 \mu -1\right)\right] z\\
   &+8 \Lambda_{\rm{L}}^5 \left[\nu +\mu  \left(2-\mu ^3-\nu  \mu ^2-\nu  \mu -\mu +\nu \right)\right],
 \end{align*}
\begin{align*}
&\chi_0=-\Lambda_{\rm{L}} \mu  (\mu +1) \left(\mu ^2-2 \mu
   -2 \nu -1\right) z^5\\
   &+\mu (\mu -1)  \left[4 \mu  (\mu +1)\Lambda_{\rm{L}}^2+4 (\mu
   +1) \nu \Lambda_{\rm{L}}^2-\mu \right] z^4\\
   &-2\Lambda_{\rm{L}} (\mu -1)\left[(\mu +1) \left(2 \mu ^2+3 \nu  \mu +\mu +\nu +1\right]\Lambda_{\rm{L}}^2-3
   \mu ^2+\mu \right) z^3 \\
   &+\Lambda_{\rm{L}}^2 \left\{\mu  \Big[\mu  (\mu  [\mu -2 \nu -17]-6 \nu
   +17)-2 \nu -11\Big]+2 \nu +2\right\} z^2 \\
   &-4\Lambda_{\rm{L}}^3 (\mu -1) \left(\mu ^3-4 \mu ^2-2
   \nu  \mu +\mu -2 \nu -2\right) z+4\Lambda_{\rm{L}}^4 (1-\mu )^2 \left(\mu ^2+\nu 
   \mu +\nu +1\right),
 \end{align*}
\begin{align*}
&c_3=-\mu ^3 z^9+\Lambda_{\rm{L}} \mu ^2
   \left(2 \mu ^2+9 \mu -1\right) z^8-2\Lambda_{\rm{L}}^2 \mu  \left[6 \mu ^3+16 \mu
   ^2-\mu +2 (\mu +1)^2 \nu +1\right] z^7 \\
   &+\Lambda_{\rm{L}} \mu  \left[2\Lambda_{\rm{L}}^2 \left(3-\mu ^4+8 \mu ^3+20 \mu ^2-2 \mu +8 (\mu +1)^2
   \nu\right)-3 \mu ^3-7 \mu^2-4 \mu \right] z^6 \\
   &+\Lambda_{\rm{L}}^2 \left\{4 \nu  \left[(\Lambda_{\rm{L}}-\Lambda_{\rm{L}} \mu )^2+\mu \right]  (\mu +1)^2\right\}z^5\\
   &+\Lambda_{\rm{L}}^2 \left\{\mu  \left[4 \left(\mu^4-2 \mu ^2+2 \mu-1\right)\Lambda_{\rm{L}}^2+\mu  (17 \mu  (\mu+2)+11)-6\right]\right\} z^5\\
   &+2\Lambda_{\rm{L}} \left\{\mu ^3+\Lambda_{\rm{L}}^2
   (\mu +1) \left[\mu  \left(\mu ^3-15 \mu ^2-11 \mu -10 (\mu +1) \nu
   +9\right)-2\right]\right\} z^4 \\
&+2\Lambda_{\rm{L}}^2 \left(-\mu ^4-6 \mu ^3+3 \mu^2\right) z^3\\
&-2\Lambda_{\rm{L}}^4 (\mu +1) \left\{2 \nu +\mu  \left[-3 \nu +\mu  (-4 \nu +\mu  (\mu +\nu-5)-2)+4\right]-2\right\}z^3 \\
   &+2\Lambda_{\rm{L}}^3  \left[5 \mu ^3+14 \mu ^2-11
   \mu -4 (\mu -1) (\mu \Lambda_{\rm{L}}+\Lambda_{\rm{L}})^2+4\right]z^2\\
  &-4\Lambda_{\rm{L}}^3\left(2\Lambda_{\rm{L}}^2 (\mu -1)-\mu \right) (\mu +1)^2 \nu  z^2\\
  &-4 \Lambda_{\rm{L}}^4 \left\{\mu  \Big[\nu +\mu  [5 \nu +\mu  (4 \mu +3 \nu +7)-7]+5\Big]-\nu -1\right\} z,
\end{align*}
and,
\begin{align*}
&\chi_1= (z-2\Lambda_{\rm{L}})^2 \mu  \left[\Lambda_{\rm{L}} \left(\mu ^2-1\right)
   z^3+\mu  z^2+2\Lambda_{\rm{L}} (1-\mu ) z-\Lambda_{\rm{L}}^2(1-\mu )^2\right]\\
   &-2\Lambda_{\rm{L}} \nu \Big[\Lambda_{\rm{L}} \left(2 z^4-3
  \Lambda_{\rm{L}} z^3-z^2+2\Lambda_{\rm{L}}^2\right) \mu ^2+\left(z^3-2
  \Lambda_{\rm{L}}\right) \left(z^2-2\Lambda_{\rm{L}} z+2 \Lambda_{\rm{L}}^2\right) \mu \\
   &+\Lambda_{\rm{L}} \left(\Lambda_{\rm{L}} z^3+z^2-4
  \Lambda_{\rm{L}} z+2\Lambda_{\rm{L}}^2\right)\Big],
\end{align*}
\begin{align*}
&c_4=\Lambda_{\rm{L}} (\mu -1)\Big[-1-2 \nu \nonumber \\
&+\mu  \left\{4 \nu  z^2+\left(4 z^2-1\right) \mu -\mu[(\mu -2) \mu +2 \nu +1] z^3-2\Lambda_{\rm{L}}^2 (\mu -1)^2 (\mu +\nu )z-2 \nu -2\right\}\Big]
\end{align*}
\begin{align*}
& c_5= \mu  \left(\mu ^2-2 \nu -1\right) z^3+2 \Lambda_{\rm{L}}^2 (\mu -1)^2 (\mu +\nu +1) z\\
&-\Lambda_{\rm{L}} (\mu -1) \left\{\mu \left[4 (\mu +\nu +1) z^2-\mu -2 \nu -2\right]-2 \nu -1\right\}.
\end{align*}
The offdiagonal right boundary vectors components \eqref{rightvecs} are
\bea
&\hspace{-20mm}\ket{\tilde{r}_{0,0}}=C_3\left( \frac{2 z \Lambda_{\rm{R}}}{1-2 z \Lambda_{\rm{R}}}c_6 , c_6 , c_7, 0 \right) 
\eea
\bea
&\hspace{-20mm}\ket{\tilde{r}_{0,1}}= C_3\left(0 , -c_6 , c_8 , 0  \right)
\eea
\bea
&\hspace{-20mm}\ket{\tilde{r}_{1,0}}=C_3 \left( 0 , \frac{c_6 (\sigma -1)}{\sigma +1} , 0 , -\frac{c_6 z \Lambda_{\rm{R}}}{\sigma +1} \right)
\eea
\bea
&\hspace{-20mm}\ket{\tilde{r}_{1,1}}=C_3\left( 0 , \frac{c_6}{1-2 z \Lambda_{\rm{R}}} , 0 , -\frac{c_6 z \Lambda_{\rm{R}}}{\sigma +1} \right)
\eea
\bea
&\hspace{-20mm}\ket{\tilde{r}_{0}}=C_4 \left( 0 , c_9 , c_{10}  , c_9 \frac{2 \Lambda_{\rm{R}}  (\rho +\sigma ) (z-2 \Lambda_{\rm{R}}) (\Lambda_{\rm{R}} (\sigma-1) z+1)}{(\Lambda_{\rm{R}} (\sigma -1)+z) (2 \Lambda_{\rm{R}}+\sigma  z)}  \right) 
\eea
\bea
&\hspace{-20mm}\ket{\tilde{r}_{1}}=C_4 \left( 0 , c_9 , c_{11}  , -c_9 \frac{2 \Lambda_{\rm{R}}  (\rho +\sigma +1)
   (z-2 \Lambda_{\rm{R}}) (\Lambda_{\rm{R}} (\sigma -1) z+1)}{(\Lambda_{\rm{R}}
   (\sigma -1)+z) (2 \Lambda_{\rm{R}}+\sigma  z)} \right)
\eea
where,
\bea
&\hspace{-15mm}C_3=\frac{\Lambda_{\rm{R}}^2 (\sigma +1) \left[2 \Lambda_{\rm{R}} \left(\rho  \sigma
   +\rho +\sigma ^2+1\right)-\sigma ^2+\sigma \right] (2 \Lambda_{\rm{R}}+\sigma 
   z)}{\sigma  z^2 \left(z^2-1\right) (z-2 \Lambda_{\rm{R}}) \left[(\sigma -1) \sigma 
   z-2 \Lambda_{\rm{R}} \left(\rho  \sigma +\rho +\sigma ^2+1\right)\right]} \nonumber\\
&\hspace{-15mm}C_4=\frac{\sigma  z^2 \left(z^2-1\right) (z-2 \Lambda_{\rm{R}}) (\Lambda_{\rm{R}}
   (\sigma -1)+z) \left[(\sigma -1) \sigma  z-2 \Lambda_{\rm{R}} \left(\rho  \sigma
   +\rho +\sigma ^2+1\right)\right]}{\Lambda_{\rm{R}}^3 (\sigma +1)^2 \left(2
   \Lambda_{\rm{R}} \left(\rho  \sigma +\rho +\sigma ^2+1\right)-\sigma ^2+\sigma
   \right)}\nonumber
\eea
\begin{align*}
&c_6= \Lambda_{\rm{R}} (\sigma +1) z (z-2 \Lambda_{\rm{R}}) (2 \Lambda_{\rm{R}} z-1),\nonumber\\
&c_7=2\Lambda_{\rm{R}} (2 \rho -(\sigma -2) \sigma +1) z^4+4 \Lambda_{\rm{R}}^2 (\sigma
   -1) (\rho +\sigma ) z^3 \\
   &-2 \Lambda_{\rm{R}} (\rho +1) (\sigma +1) z^2+[(\sigma -2) \sigma -2 \rho-1] z+2 \Lambda_{\rm{R}} [2 \rho -(\sigma -2) \sigma +1],\\
&c_8= 2 \Lambda_{\rm{R}} \left(\sigma ^2-2 \rho -1\right) z^4- \\
    &4\Lambda_{\rm{R}}^2 (\sigma -1) (\rho +\sigma +1) z^3+2 \Lambda_{\rm{R}} \rho 
   (\sigma +1) z^2+\left(1-\sigma ^2+2 \rho\right) z+2 \Lambda_{\rm{R}} \left(\sigma
   ^2-2 \rho -1\right),\\
&c_9=\Lambda_{\rm{R}} z^2 (z-2 \Lambda_{\rm{R}}) (\Lambda_{\rm{R}} (\sigma -1)+z) (2\Lambda_{\rm{R}}+\sigma  z),\nonumber\\
&c_{10}=-2 \Lambda_{\rm{R}}^2 (\sigma -1) (\sigma +1)^2-(\sigma -1) \sigma 
   z^6+4 \Lambda_{\rm{R}} (\sigma -1) \sigma  z^5+2 \Lambda_{\rm{R}}^2 (\sigma -1)z^4 \left[(\rho -1) \sigma +\rho +\sigma ^2\right]\\
   &+\Lambda_{\rm{R}} z^3 \left\{2
   \rho  \left[\sigma-2 \Lambda_{\rm{R}}^2 \left(\sigma ^2-1\right) +1\right]+\sigma 
   \left[1-4 \Lambda_{\rm{R}}^2 \left(\sigma ^2-1\right)-(\sigma -2) \sigma
   \right]+2\right\}\\
   &-4 \Lambda_{\rm{R}}^2 z^2 \left(\rho  \sigma +\rho -\sigma
   ^3+\sigma ^2+2\right)+4 \Lambda_{\rm{R}}^3 (\sigma -1)^2 z,\\
&c_{11}= (\sigma +1)\Big(\sigma  z^6+4 \Lambda_{\rm{R}}^3 (\sigma -1) z
   \left[z^2 (\rho +\sigma +1)-1\right]   -\Lambda_{\rm{R}} z^3 \left[2 \rho +\sigma  \left(1-\sigma +4
   z^2\right)\right]\\
   &+2 \Lambda_{\rm{R}}^2 \left\{(\sigma +1)^2+z^4
   [1-\rho \sigma+\rho -(\sigma -2) \sigma]+2 z^2 \left(\rho -\sigma^2-1\right)\right\}
 \Big).
\end{align*}


\section*{References}

\begin{thebibliography}{10}
\bibitem{review} T. Prosen, Topical Review:  {\em Matrix product solutions of boundary driven quantum chains},  J. Phys. A: Math. Theor. 48, 373001 (2015).
\bibitem{drivenbobenko} T. Prosen and C. Mej\'ia-Monasterio, {\em Integrability of a deterministic cellular automaton driven by stochastic boundaries}, J. Phys. A: Math. Theor. {\bf 49} 185003 (2016).

\bibitem{Bobenko} A. Bobenko, M. Bordermann, C. Gunn, U. Pinkall, {\em On Two Integrable Cellular Automata}, Commun. Math. Phys. {\bf 158}, 127 (1993).

\bibitem{asep1} G. Sch\"utz, {\em Time-dependent correlation functions in a one-dimensional asymmetric exclusion process}, Phys. Rev. E {\bf 47} 4265 (1993);
H. Hinrichsen, {\em Matrix product ground states for exclusion processes with parallel dynamics}, J. Phys. A: Math. Gen. {\bf 29} 3659 (1996);
N. Rajewsky, L. Santen, M. Schreckenberg and A. Schadschneide, {\em The asymmetric exclusion process: Comparison of update procedures}, J. Stat. Phys. {\bf 92} 151 (1998).

\bibitem{asep2} M. R. Evans, N. Rajewsky, and E. R. Speer, {\em Exact solution of a cellular automaton for traffic},J. Stat. Phys {\bf 95} 45 (1999); 
J. Brankov, N. Pesheva and N. Valkov, {\em Exact results for a fully asymmetric exclusion process with sequential dynamics and open boundaries}, Phys. Rev. E {\bf 61} 2300–2318 (2000); 
D. Chowdhury, L. Santen, A. and Schadschneider, Statistical physics of vehicular traffic and some related systems Physics Reports {\bf 329} 199 (2000);
J. de Gier and B. Nienhuis, {\em Exact stationary state for an asymmetric exclusion process with fully parallel dynamics},
Phys. Rev. E {\bf 59} 4899 (1999)

\bibitem{asep3} R. A. Blythe and M. R. Evans, {\em Nonequilibrium steady states of matrix-product form: a solver's guide}, Phys. A: Math. Theor. {\bf 40} R333 (2007).

\bibitem{matrixansatzASEP} N. Crampe, E. Ragoucy and D. Simon, {\em Matrix coordinate Bethe Ansatz: applications to XXZ and ASEP models}, J. Phys. A: Math. Theor. {\bf 44} 405003 (2011).  

\bibitem{reacmodel} R. A. Blythe, M. R. Evans, and Y. Kafri, {\em Stochastic Ballistic Annihilation and Coalescence}, Phys. Rev. Lett. {\bf 85}, 3750 (2000). 

\bibitem{intCA1} T. Tokihiro, D. Takahashi, J. Matsukidaira, and J. Satsuma, {\em From Soliton Equations to Integrable Cellular Automata through a Limiting Procedure}, Phys. Rev. Lett. {\bf 76}, 3247 (1996).

\bibitem{intCA2} M. Bruschi, and P.~M. Santini, {\em Cellular automata in 1 + 1, 2 + 1 and 3 + 1 dimensions, constants of motion and coherent structures}, Physica D {\bf 70}, 185  (1994).

\bibitem{intCA3} N. Joshi, S. Lafortune, {\em How to detect integrability in cellular automata}, J. Phys. A: Math. Gen. {\bf 38} L499 (2005). 

\bibitem{intCA4} L. G. Tilstra and M. H. Ernst, {\em Synchronous asymmetric exclusion processes}, J. Phys. A: Math. Gen. {\bf 31} 5033 (1998). 

\bibitem{YF} A. Yu. Volkov and L. D. Faddeev, {\em Quantum inverse scattering method on a spacetime lattice}, Theor. Math. Phys. {\bf 92}: 837 (1992). 
 
\end{thebibliography}
\end{document}